\documentclass{article}[12pt]
\usepackage[utf8]{inputenc}
\usepackage{subfiles}
\usepackage{array} 
\newtheorem{theorem}{Theorem}

\newtheorem{remark}[theorem]{Remark}

\usepackage{graphicx}
\usepackage{epstopdf}
\usepackage{xcolor}
\newcommand{\mgray}[1]{\textcolor{gray}{#1}}
\definecolor{grey}{RGB}{128,128,128} 
\usepackage{setspace}
\usepackage{fancyhdr}
\usepackage{lscape}
\usepackage{longtable}
\usepackage{threeparttable}
\usepackage{booktabs}
\usepackage{pgfplots}
\usepackage{amsmath}
\usepackage{quoting}
\usepackage{makecell}
\usepackage{graphicx}
\usepackage{siunitx}

\usepackage{natbib}
\usepackage[size=normalsize, font=bf]{caption}
\usepackage{amsmath}
\usepackage{float} 
\usepackage{url}
\usepackage{placeins}
\usepackage{subfigure}
\usepackage{subcaption}
\usepackage[bottom]{footmisc}
\usepackage{hyperref}
\usepackage{tikz}
\usepackage[margin=1.2in]{geometry}
\usepackage{pgfplots}
\usepackage{xr}
\pgfplotsset{compat=1.17}

\definecolor{crimson}{RGB}{156,0,0}
\hypersetup{colorlinks = true, citecolor = black, linkcolor = crimson, urlcolor = black, plainpages = false}

\date{}

\newcommand{\tikzpicone}{
    \begin{tikzpicture}
        \coordinate (x) at (0,0);
        \coordinate (x1) at (3,0);
        \coordinate (y) at (0,-3);
        \coordinate (lin1) at (0.2,-0.2);
        \coordinate (lin2) at (2.8,-2.8);

  \draw[->] (x) -- (y) node[anchor=north] {$u_i(\delta_{i,c})$};
    \draw[->] (x) -- (x1) node[anchor=west] {$\delta_{i,c}$};
        \draw (lin1) -- (lin2);

        \foreach \point in {}
            \fill (\point) circle (3pt) node[below] {\point};
    \end{tikzpicture}
}

\newcommand{\tikzpictwo}{ 
\begin{tikzpicture}
     \coordinate (x) at (0,0);
    \coordinate (x1) at (4,0);
    \coordinate (y) at (0,4);
    \coordinate (lin1) at (0.2,-0.2);
    \coordinate (lin2) at (2.8,-2.8);

    \draw[->] (x) -- (y) node[anchor=south] {$x_2$};
    \draw[->] (x) -- (x1) node[anchor=west] {$x_1$};
    \draw (3.6,4) arc (180:270:0.4cm);
    \draw (2.3,4) arc (180:270:1.7cm);
    \draw (1.5,4) arc (180:270:2.5cm);
    \draw (0.95,4) arc (180:270:3.05cm);
    \draw (0.6,4) arc (180:270:3.4cm);
     \draw (0.4,4) arc (180:270:3.6cm);

    \foreach \point in {}
        \fill (\point) circle (2pt) node[below] {\point};
\end{tikzpicture}
}

\newcommand{\tikzpicthree}{ 
 \begin{tikzpicture}
     \coordinate (x) at (0,0);
    \coordinate (x1) at (4,0);
    \coordinate (y) at (0,4);
    \coordinate (lin1) at (0.2,-0.2);
    \coordinate (lin2) at (2.8,-2.8);

    \draw[->] (x) -- (y) node[anchor=south] {$x_2$};
    \draw[->] (x) -- (x1) node[anchor=west] {$x_1$};
    \draw (3.6,4) arc (180:270:0.4cm);
    \draw (3.4,4) arc (180:270:0.6cm);
    \draw (3.05,4) arc (180:270:0.95cm);
    \draw (2.5,4) arc (180:270:1.5cm);
    \draw (1.7,4) arc (180:270:2.3cm);
     \draw (0.4,4) arc (180:270:3.6cm);

    \foreach \point in {}
        \fill (\point) circle (2pt) node[below] {\point};
\end{tikzpicture}
}

\newcommand{\tikzpicfour}{ 

 \begin{tikzpicture}
     \coordinate (x) at (0,0);
    \coordinate (x1) at (4,0);
    \coordinate (y) at (0,4);
    \coordinate (lin1) at (0.2,-0.2);
    \coordinate (lin2) at (2.8,-2.8);

    \draw[->] (x) -- (y) node[anchor=south] {$x_2$};
    \draw[->] (x) -- (x1) node[anchor=west] {$x_1$};
    \draw (3.6,4) arc (180:270:0.4cm);
    \draw (2.8,4) arc (180:270:1.2cm);
    \draw (2.4,4) arc (180:270:1.6cm);
    \draw (2.2,4) arc (180:270:1.8cm);
    \draw (1.7,4) arc (180:270:2.3cm);
     \draw (0.4,4) arc (180:270:3.6cm);

    \foreach \point in {}
        \fill (\point) circle (2pt) node[below] {\point};
\end{tikzpicture}
}

\newcommand{\tikzpicsix}{ 
\begin{tikzpicture}
     \coordinate (x) at (0,0);
    \coordinate (x1) at (4,0);
    \coordinate (y) at (0,4);
    \coordinate (lin1) at (0.2,-0.2);
    \coordinate (lin2) at (2.8,-2.8);

    \draw[->] (x) -- (y) node[anchor=south] {$x_2$};
    \draw[->] (x) -- (x1) node[anchor=west] {$x_1$};
    \draw (3.6,4) arc (180:270:0.4cm);
    \draw (2.3,4) arc (180:270:1.7cm);
    \draw (1.5,4) arc (180:270:2.5cm);
    \draw (0.95,4) arc (180:270:3.05cm);
    \draw (0.6,4) arc (180:270:3.4cm);
     \draw (0.4,4) arc (180:270:3.6cm);

    \foreach \point in {}
        \fill (\point) circle (2pt) node[below] {\point};
\end{tikzpicture}
}

\newcommand{\tikzpicseven}{ 
\begin{tikzpicture}
     \coordinate (x) at (0,0);
    \coordinate (x1) at (4,0);
    \coordinate (y) at (0,4);
    \coordinate (lin1) at (0.2,-0.2);
    \coordinate (lin2) at (2.8,-2.8);

    \draw[->] (x) -- (y) node[anchor=south] {$x_2$};
    \draw[->] (x) -- (x1) node[anchor=west] {$x_1$};

\draw(3.5,4) ..controls (3.8,3.8) .. (4,3.5);
\draw(3.05,4) ..controls (3.65,3.65) .. (4,3.05);
\draw(2.5,4) ..controls (3.45,3.45) .. (4,2.5);
\draw(1.7,4) ..controls (3.21,3.21) .. (4,1.7);
\draw(1,3.8) ..controls (2.85,2.85) .. (3.8,1);
\draw(0.4,3.25) ..controls (2.25,2.25) .. (3.25,0.4);
\draw(0.4,2) ..controls (1.45,1.45) .. (2,0.4);

    \foreach \point in {}
        \fill (\point) circle (2pt) node[below] {\point};
\end{tikzpicture}
} 

\newcommand{\tikzpiceight}{ 
\begin{tikzpicture}
     \coordinate (x) at (0,0);
    \coordinate (x1) at (4,0);
    \coordinate (y) at (0,4);
    \coordinate (lin1) at (0.2,-0.2);
    \coordinate (lin2) at (2.8,-2.8);

    \draw[->] (x) -- (y) node[anchor=south] {$x_2$};
    \draw[->] (x) -- (x1) node[anchor=west] {$x_1$};

\draw(3.5,4) ..controls (3.65,3.65) .. (4,3.5);
\draw(2.5,4) ..controls (3.1,3.1) .. (4,2.5);
\draw(1.85,4) ..controls (2.8,2.8) .. (4,1.85);
\draw(1.5,4) ..controls (2.75,2.75) .. (4,1.5);
\draw(1.3,3.9) ..controls (2.72,2.72) .. (3.9,1.3);
\draw(1.1,3.7) ..controls (2.65,2.65) .. (3.7,1.1);
\draw(0.4,3.25) ..controls (2.25,2.25) .. (3.25,0.4);
\draw(0.4,2) ..controls (1.45,1.45) .. (2,0.4);

    \foreach \point in {}
        \fill (\point) circle (2pt) node[below] {\point};
\end{tikzpicture}
}

\newcommand{\tikzpicsecond}{
    \begin{tikzpicture}
        \coordinate (x) at (0,0);
        \coordinate (x1) at (3,0);
        \coordinate (y) at (0,-3);
        \coordinate (lin1) at (0.2,-0.2);
        \coordinate (lin2) at (2.8,-2.8);

    \draw[->] (x) -- (y) node[anchor=north] {$u_i(\delta_{i,c})$};
    \draw[->] (x) -- (x1) node[anchor=west] {$\delta_{i,c}$};
        \draw (0.2,-0.2) parabola (2.8,-2.8);

        \foreach \point in {}
            \fill (\point) circle (2pt) node[below] {\point};
    \end{tikzpicture}
}

\newcommand{\tikzpicthird}{
    \begin{tikzpicture}
        \coordinate (x) at (0,0);
        \coordinate (x1) at (3,0);
        \coordinate (y) at (0,-3);
        \coordinate (lin1) at (0.2,-0.2);
        \coordinate (lin2) at (2.8,-2.8);
  
    \draw[->] (x) -- (y) node[anchor=north] {$u_i(\delta_{i,c})$};
    \draw[->] (x) -- (x1) node[anchor=west] {$\delta_{i,c}$};
        \draw (2.8,-2.8) parabola (0.2,-0.2);

        \foreach \point in {}
            \fill (\point) circle (2pt) node[below] {\point};
    \end{tikzpicture}
}

\newcommand{\tikzpicfourth}{
    \begin{tikzpicture}
        \coordinate (x) at (0,0);
        \coordinate (x1) at (3,0);
        \coordinate (y) at (0,-3);
        \coordinate (lin1) at (0.2,-0.2);
        \coordinate (lin2) at (2.8,-2.8);

         \draw[->] (x) -- (y) node[anchor=north] {$u_i(\delta_{i,c})$};
          \draw[->] (x) -- (x1) node[anchor=west] {$\delta_{i,c}$};
        \draw (0.2,-0.2) parabola (1.4,-1.4);
        \draw (2.8,-2.8) parabola (1.4,-1.4);

        \foreach \point in {}
            \fill (\point) circle (2pt) node[below] {\point};
    \end{tikzpicture}
}

\newcommand{\tikzindiffoverKone}{
    \begin{tikzpicture}
    \coordinate (x) at (0,0);
    \coordinate (x1) at (4,0);
    \coordinate (y) at (0,-3);
    \coordinate (lin1) at (0.2,0);
    \coordinate (lin2) at (3.8,0);

    \draw[->] (x) -- (y)   node[anchor=north] {$-|u(\delta_{i,c_{D1}})-u(\delta_{i,c_{D2}})|$};
    \draw[ ->] (x) -- (x1) node[anchor=west] {$x_i$};
    \draw [ultra 
    thick] (lin1) -- (lin2);

    \foreach \point in {}
        \fill (\point) circle (2pt) node[below] {\point};
\end{tikzpicture}
}

\newcommand{\tikzindiffoverKtwo}{

    \begin{tikzpicture}
     \coordinate (x) at (0,0);
    \coordinate (x1) at (4,0);
    \coordinate (y) at (0,-3);
    \coordinate (lin1) at (0.2,-0.2);
    \coordinate (lin2) at (2.8,-2.8);

    \draw[->] (x) -- (y) node[anchor=north] {$-|u(\delta_{i,c_{D1}})-u(\delta_{i,c_{D2}})|$};
    \draw[->] (x) -- (x1) node[anchor=west] {$x_i$};
    \draw (0.2,-1) parabola (3.8,-2.8) ;

    \foreach \point in {}
        \fill (\point) circle (2pt) node[below] {\point};
\end{tikzpicture}
}
\newcommand{\tikzindiffoverKthree}{

\begin{tikzpicture}
     \coordinate (x) at (0,0);
    \coordinate (x1) at (4,0);
    \coordinate (y) at (0,-3);
    \coordinate (lin1) at (0.2,-0.2);
    \coordinate (lin2) at (2.8,-2.8);

    \draw[->] (x) -- (y) node[anchor=north] {$-|u(\delta_{i,c_{D1}})-u(\delta_{i,c_{D2}})|$};
    \draw[->] (x) -- (x1) node[anchor=west] {$x_i$};
    \draw (3.8,-0.2) parabola (0.2,-2.8) ;

    \foreach \point in {}
        \fill (\point) circle (2pt) node[below] {\point};
\end{tikzpicture}
}

\newcommand{\tikzindiffoverKfour}{

    \begin{tikzpicture}
     \coordinate (x) at (0,0);
    \coordinate (x1) at (4,0);
    \coordinate (y) at (0,-3);
    \coordinate (lin1) at (0.2,-0.2);
    \coordinate (lin2) at (2.8,-2.8);

    \draw[->] (x) -- (y) node[anchor=north] {$-|u(\delta_{i,c_{D1}})-u(\delta_{i,c_{D2}})|$};
    \draw[->] (x) -- (x1) node[anchor=west] {$x_i$};
    \draw  (1,-2.8) parabola (2.4,-1.5);
    \draw (1,-2.8) parabola (0.4,-2.5);
    \draw  (0.2,-2.4) parabola  (0.4,-2.5);
    \draw (3.8,-0.2) parabola (2.4,-1.5);

    \foreach \point in {}
        \fill (\point) circle (2pt) node[below] {\point};
\end{tikzpicture}
}

\newcommand{\tikzpicfive}{ 
 \begin{tikzpicture}
     \coordinate (x) at (0,0);
     \coordinate (x1) at (4,0);
     \coordinate (y) at (0,4);
     \coordinate (lin1) at (0.2,-0.2);
     \coordinate (lin2) at (2.8,-2.8);

     \draw[->] (x) -- (y) node[anchor=south] {$x_2$};
     \draw[->] (x) -- (x1) node[anchor=west] {$x_1$};
     \draw (3.6,4) -- (4,3.6); 
     \draw (2.8,4) -- (4,2.8); 
     \draw (2.0,4) -- (4,2);   
     \draw (1.2,4) -- (4,1.2); 
     \draw (0.4,4) -- (4,0.4); 
     \draw (0.4,3.2) -- (3.2,0.4); 

     \foreach \point in {}
         \fill (\point) circle (2pt) node[below] {\point};
 \end{tikzpicture}
}

\begin{document}

\begin{titlepage}
\title{\Large {\bf When Do Voters Stop Caring? 

Estimating the Shape of Voters' Utility Functions}\footnote{We owe great thanks to Jim Snyder, without whom this paper could never have been possible, and to Stephane Wolton for his significant input and continued guidance. We also thank Peter Buisseret, Maggie Penn, Rocio Titiunik, Jeffry Frieden, Justin Melnick, and participants of EITM 2023 for their helpful comments.}}
 
\author{
\large Aleksandra Conevska\footnote{PhD candidate at the Harvard University Government Department. Email: aleksandraconevska@g.harvard.edu } \\
\large Can Mutlu \footnote{PhD candidate at the Harvard University Government Department. Email: cmutlu@g.harvard.edu}\\
}

\date{\normalsize January 2025}

\clearpage
\maketitle
\thispagestyle{empty}

\begin{abstract}
\noindent
\singlespacing
\normalsize
In this paper, we address a longstanding puzzle over the functional form that better approximates voter utility from political choices. Though it has become the norm in the literature to represent voter utility with concave loss functions, for decades scholars have underscored this assumption's potential shortcomings. Yet there exists little to no evidence to support one functional form assumption over another. We fill this gap by first identifying electoral settings where the different functional forms generate divergent predictions about voter behavior. We then assess which functional form better matches observed voter and abstention behavior using Cast Vote Record (CVR) data that captures the anonymized ballots of millions of voters in the $2020$ U.S. general election. Our findings indicate that concave loss functions fail to predict voting and abstention behavior, and it is the reverse S-shaped loss functions, such as the Gaussian function, that better match observed voter behavior. 
\end{abstract}

\end{titlepage}

\section*{Introduction}


All theoretical and empirical models of electoral politics are built explicitly or implicitly on some assumptions about the preferences of voters. Specifically, how voters respond to candidate choices and their positions is captured through the functional form of voter's utility. The dominant assumption made in the literature is that the voter utility over candidate choices is concave, often in the form of single-peaked quadratic or linear loss function for analytical convenience. Yet, are these assumptions over the voter utility's functional form empirically valid? There exists little to no evidence to support one functional form, whether it is strictly concave, strictly convex, linear or reverse S-shaped, over the other \citep{singh2014linear}. In this paper, we use unique individual level data from the 2020 U.S. election to identify electoral races where these assumptions yield divergent predictions, and empirically test which functional form best predicts observed voter behavior. We find little support for the widely accepted concave loss functions and instead demonstrate that the reverse S-shaped function best predicts observed voter behavior. We then consider the implications of these findings for the theoretical and empirical models of electoral politics and how our findings can help explain political phenomena that we are yet to explain.

The lack of existing empirical support as to which functional form best describes observed behavior is important because depending on the assumption made regarding the shape of voter utility, we obtain substantively different predictions over aspects of political behavior that are significant to the study of politics. This includes, though certainly is not limited to, (1) party positioning in theoretical models and whether they converge in equilibrium; (2) which voters we can expect to vote and when, and when we can expect them to abstain; and (3) when and how voters should be especially likely to vote strategically, and when not, particularly in multi-candidate races. Our understanding of these questions is effectively constrained by our knowledge of the true underlying shape of voter utility. 

The different functional form assumptions yield divergent predictions over expected behavior because each generates different implications specifically regarding when voters are going to be indifferent, or approximately indifferent. Put another way, since the functional form of voter utility captures how sensitive voters are to differences between candidate positions, it dictates when voters are going to care a lot, or not at all. To see this, consider a very simple example about the widely accepted concave loss function for a Bernie Sanders supporter. When this voter faces a choice between voting in the Democratic Party primary between Bernie Sanders and Hillary Clinton or the Republican Party primary between Donald Trump and Jeb Bush, should we expect the voter to participate in the Democratic or Republican primary? With concave voter utility, we would predict that this Bernie supporter will vote in the Republican primary, as their utility is more affected by the differences between the Republican candidates. These examples are present across various political contexts. If we assume that voter utility is concave, this implies that voters on the extreme ends of the ideological spectrum will be highly sensitive to differences between two or more moderate candidates. But do we have any reason to believe that a MAGA voter's utility will be \emph{highly} sensitive to a choice between say, Lisa Murkowski and Joe Manchin? For MAGA voters to place a much higher value on one of these candidates over the other, they must perceive a great difference between them. Our findings allows\ us to contextualize and better explain these empirical observations from modern politics. 

To empirically test the divergent functional form predictions, we require a measure of distance between the voters' ideal policy preferences and the candidate positions, as well as a measure of voters' indifference to candidate choices. We measure voter distance to the Democratic and Republican Party candidates using state-level ballot measures. With our CVR data for millions of individuals, and with the political parties' declared positions on these measures, we group voters ordinally based on their relative distances to the political parties. We create this \emph{Voter Group} measure for the two states in our sample with the greatest number of state-wide ballot measures in $2020$: California and Colorado, allowing us to place voters and candidates on a single policy space in each state. We use two measures to capture voter indifference: abstention and vote predictability. The average rate of abstention for each \emph{Voter Group} and the net swing towards the \emph{Voter Group}'s preferred candidate (our vote predictability measure), are negatively correlated, as we would expect them to be if they capture the indifference phenomenon. Importantly, our observational setting is such that voters have already turned out to vote. 

We identify two key empirical settings where the functional forms diverge in their predictions over voter indifference levels across candidate choices. In the first, which we call \emph{Case 1}, we focus on electoral races where both candidates on the ballot are from the same party competing in the general election, which we observe in our sample thanks to the pooled primary system in California. We examine how the unique \emph{Voter Group}s vary in their indifference levels in these races, and observe that as voters get more conservative (i.e. move further away from the Democratic Party position on the state measures), they grow more indifferent to the candidates, abstaining at significantly higher rates and having lower vote predictability. Importantly, the reverse trend holds true in races between two Republican candidates.

For the second electoral setting, or \emph{Case 2}, we focus on the voting behavior of specific a \emph{Voter Group}, who we call moderate voters. Voters in the moderate \emph{Voter Group}(s) exhibit the largest level of indifference to opposing party candidates across partisan office elections. Fixing the \emph{Voter Group}s of interest, we use the varying levels of candidate polarization in electoral races between the Democratic and Republican candidates in California and Colorado. Here, we observe that as the opposing party candidates get more polarized, first, the moderate voters become less indifferent to the candidates and then grow more indifferent as the level of candidate polarization further increases. In both cases, the predictions of the reverse S-shaped function best support the observed voter behavior. We believe our evidence can unify the approaches taken by the formal and empirical literature, where the reverse S-shaped function is already used in the ideal point estimation model NOMINATE and in a limited number of formal theoretical models \citep{poole1985spatial, poole2008scaling, callander2007turnout}.

Our evidence has significant implications for both theoretical and empirical models of voter behavior and electoral competition. For one, it provides a much clearer explanation for the abstention behavior of different voter groups by unifying existing concepts of abstention due to indifference versus that due to alienation. Our empirical support for the reverse S-shaped loss function is also consequential even for the most fundamental results of median voter theorem in probabilistic voting models, and in deterministic voting models when abstention is allowed. It also provides a basis to explain one of the earliest intuitions in the quantitative studies of politics: \textit{The possibility that parties will be kept from converging ideologically in a two-party system depends upon the refusal of extremist voters to support either party if both become alike– not identical, but merely similar} \citep{downs1957jpe}. In these settings, the equilibrium results for candidates' strategic policy positions will depend not on the ideal policy position of the median voter but the distribution of voters' ideal positions leading to divergence from the median voter's ideal position in equilibrium when voters are sufficiently polarized. It is also possible to have cases where no pure strategy equilibrium exists even when the policy space is taken to be uni-dimensional.

Our findings can also help us better understand the effect on valence shocks on different voter groups as well as the parties', candidates' and interest groups' rationale for targeting campaign spending and vote buying activities on certain voter groups. Finally, in addition to the ideal point estimations in the empirical literature, our result enables us to reassess the power of linear regression models that aim to discern the effect of issue positions on electoral behavior and outcomes. Our evidence indicates that voter responsiveness to candidate positions as a function of their preference on the issue will not necessarily follow a monotone trend, and the impact of certain policy issues on electoral outcomes might be better understood using non-linear regression models that better reflect the predictions of the reverse S-shaped functional form.

\section*{Background}

In the formal theoretical literature, scholars have focused on the implications of functional form assumptions on turnout and candidates' strategic platform choices. Although concave loss functions have become the norm, mainly for mathematical convenience, early as well as contemporary scholars have raised qualms about it's validity, and have investigated the implications of other functional form assumptions, especially when modeling turnout behavior. As Callander and Wilson (2007) summarize: 
\begin{quote}
Early formal modelers also demonstrated enthusiasm to incorporate abstention into their work, in particular into models of electoral competition. Indeed, one of the first extensions offered to the pioneering spatial model of Hotelling (1929) was the inclusion of abstention (Smithies 1941, see also Downs 1957). This enthusiasm did not last long, due in part to the findings of initial studies (Hinich and Ordeshook 1969; Ledyard 1984) that abstention makes little difference to the behavior of strategic candidates. In recent decades, consequently, most prominent formal models of electoral competition have begun with the empirically unrealistic assumption of full voter turnout (Besley and Coate, 1997; Calvert, 1985; Osborne and Slivinski, 1996; Palfrey, 1984; Wittman, 1983). 
\end{quote}

The assumption regarding full turnout eliminated some concerns over the functional form of voter utility in deterministic models of electoral politics and allowed the concave loss function to be used without loss of generality. Yet when abstention is accounted for, or a probabilistic voting model is adopted, it can be shown that ''the standard assumption of concave utility is not without loss of generality'' \citep{valasek2012turnout, kamada2014voter}. Contrary to the general perception that formal theorists are distant from empirical reality, since the early days of this literature, scholars, including Downs, McKelvey, and Osborne, have raised considerable doubt as to whether the concave loss function accurately describes voter behavior \citep{downs1957jpe, mckelvey1976symm, osborne1995spatial}. Osborne (1995) notes, ``Concavity is often assumed. However, I am uncomfortable with the implication of concavity that extremists are highly sensitive to differences between moderate candidates. I conclude that in the absence of any convincing empirical evidence, it is not clear which of the assumptions is more appropriate.''

These assumptions over the functional form of voter utility are also implicitly at play in empirical research on voting behavior. For instance, a linear regression model investigating the effect of specific issues on different ideological voter groups' turnout decisions and candidate choices implicitly relies on a linear or quadratic loss function over voter utility. As we discuss in further detail below, these assumptions can lead to inaccurate estimations over the predictive significance of the issue and the voters' true responses to the issue. 

Similarly,  ideal point estimators - which generate voter, or more commonly legislator, ideal points from their observed voting behavior (e.g. roll call votes) -  rely heavily on the functional form assumption that is made over voter loss. Such models have been applied to U.S. Congress \citep{poole1985spatial, poole2008scaling}, the European Union (EU) Council of Ministers \citep{hagemann2007applying}, firm behavior \citep{ichniowski1997}, and even courts \citep{martin2002dynamic}. The most widely adopted ideal point estimators are Poole and Rosenthal's NOMINATE, which assumes a Gaussian utility model, and Jackman's IDEAL (2004), which is based on a quadratic utility model. \citet{carroll2013structure}, acknowledging the significance of the functional form assumption for NOMINATE and IDEAL, investigate their validity empirically and show that the reverse S-shaped loss functions best approximate observed legislator behavior. Though relevant to our goal, we cannot be confident that the loss function that best describes legislator behavior generalizes to voters. 

On the voter side, some empirical work has attempted to show that turnout behavior is largely affected by demographics \citep{geys2006explaining, timpone1998structure},  lack of individual level observational data has limited the ability for researchers to establish convincing evidence that voter preferences, rather than other factors, such as cost of voting, drive these results.


\section*{Formal Predictions}
\label{sec:form_pred}

In this section, we outline the formal theoretical implications of the functional form assumptions and identify cases where their predictions diverge. Of the approaches to formulating the voter's utility function, the spatial approach has been the most widely adopted in the formal and empirical literature.\footnote{Beginning with Hotelling's model of spatial competition, the canonical theoretical results of several theorems are also built on this framework: the median voter theorem, ``chaos'' theorems, citizen-candidate models, and probabilistic voting models, and the ideal point estimators that are widely used in the empirical literature, such as NOMINATE and IDEAL (\citep{poole1985spatial}, \citep{clinton2004stati}).} Here we provide the clearest demonstration of our evidence by focusing on the uni-dimensional spatial model. In the 
\hyperref[sec:append]{Appendix}, we demonstrate that our findings hold consistently in the multi-dimensional setting, and discuss their implications for other approaches to formulating voter utilities such as the directional and issue based models of voting.\footnote{Issue based models rest on the assertion that voters evaluate their utility from each policy dimension separately. The directional approach, on the other hand, is built on the assertion that voters do not necessarily prefer candidates whose policy positions best match their own, but prefer candidates with more extreme positions if the directionality of the voter's and the candidate's preferences align, and more moderate candidates, otherwise. Our analysis is based entirely on the spatial approach, but we later discuss our findings' implications for the issue based and directional approaches.} We approach our question by fixing, or controlling for, the formulation of voter utilities to then test the predictions of the functional forms.     

Consider a uni-dimensional policy space represented in a line segment $x$, along which two candidates, $c_1$ and $c_2$, take policy positions and a representative voter $i$ has an ideal policy position. Further, denote the ideal policy position of the representative voter on the policy space by $x_i$ and the policy positions of the two candidates by $x_{c_1}$ and and $x_{c_2}$, respectively. Then, a voter's utility from each candidate is a function of the Eucledian distance between the voter's ideal position and the positions of the candidate. Formally, a voter's utility from the candidates $c_1$ and $c_2$ is expressed as $u(\delta_{i,c_1})$ and $u(\delta_{i,c_2})$, where $\delta_{i,c_1}=|x_i-x_{c_1}|$ and $\delta_{i,c_2}=|x_i-x_{c_2}|$ are the spatial distances between the voter's ideal policy position and the respective candidate's position. While $u(.)$ is a monotone loss function that is single peaked at $\delta=0$ and decreasing in spatial distances, $\frac{d u(\delta)}{d \delta}<0$, the crucial question at hand is over the functional form of the loss function $u(.)$.

The specific form of the loss function, whether it is linear, strictly concave, strictly convex, or reverse S-shaped, reflects how sensitive the voter's utility is to the spatial distances. Formally, this corresponds to variation in the term $\frac{d^2 u(\delta)}{d \delta^{2}}$. The mathematical properties and representative shapes of the linear, strictly concave, strictly convex, and reverse-sigmoid loss functions are displayed in Table \ref{table_func_forms}. 

Voter $i$'s candidate choice is deemed to be a function of their net utility from that candidate which we also refer to as the stakes at the election. In the context of voting in a race between candidates $c_1$ and $c_2$, voter $i$ will be expected to vote for candidate $c_1$ if she receives a sufficiently large net utility from this candidate, $u(\delta_{i,c_1})-u(\delta_{i,c_2})>c$, and vote for candidate $c_2$ if $u(\delta_{i,c_2})-u(\delta_{i,c_1})>c$, where $c$ is interpreted as the cost of voting, acquiring information about the candidates, or a psychological threshold for casting a vote in the race. In models in which voter turnout and participation is imposed, this cost is $c=0$. If, on the other hand, the voter's net utility from either candidate does not surpasses this threshold such that the stakes at the election are not sufficiently large, $|u(\delta_{i,c_2})-u(\delta_{i,c_1})|<c$, the voter is expected to abstain in the race. Hence, voter indifference can be captured by the term $-|u(\delta_{i,c_1})-u(\delta_{i,c_2})|$, where higher values represent growing voter indifference between the candidate choices.  

Given the limitations in measuring the ideal policy positions of voters and respective candidate platforms, spatial models are usually complemented with a probabilistic component. In models of probabilistic voting, the probability that voter $i$ votes for candidate $c_1$ and $c_2$ can be expressed as $Pr_i(c_1)= P(u(\delta_{i,c_1})-u(\delta_{i,c_2})-c)$ and $Pr_i(c_2)= P(u(\delta_{i,c_2})-u(\delta_{i,c_1})-c)$, where $P(.)$ is the cumulative distribution function over a single peaked distribution function, such as the unit normal distribution in the reverse S-shaped probabilistic voting models (e.g. logit and probit models). In this framework, voter $i$ abstains in a race with probability $Pr_i(A)=1-Pr_i(c_1)-Pr_i(c_2)$. Realize that in these models, the probability of abstaining in a race also increases in the term $-|u(\delta_{i,c_1})-u(\delta_{i,c_2})|$, which captures the voter's indifference to the candidates. Then, how well the probabilistic voting model predicts voter $i$'s vote for one candidate or the other is given by $|Pr_i(c_1)-Pr_i(c_2)|$, a measure of the overall predictability of an individual voter's vote in a given race, which is negatively correlated with the probability of abstaining. 

Other approaches to modeling abstention also exist. The most prominent alternative, which we refer to as the alienation approach, views abstention not as a function of the voter's net utility from the candidates or stakes in the election but the utility they receive from their preferred candidate - i.e. the candidate from which they receive a higher utility. Under this formulation, the probability of abstention is a monotone decreasing function of $\max\{u(\delta_{i,c_1}),u(\delta_{i,c_2})\}$. Though in some cases this approach bears similar implications to ours, where we model abstention as a function of the voter's indifference to the candidates, it also has counterintuitive implications. As a stark example, consider a liberal voter who has the choice to either vote or abstain in a race between two candidates, Joe Biden and Joe Biden. Now consider the same liberal voter's abstention decision in a race between Joe Biden and Donald Trump. Under the alienation approach, this voter is equally as likely to abstain in each of these two races. Further, in \emph{Case 2}, we show that the relationship between alienation and abstention is not monotone as suggested by the alienation approach. We discuss other implications and fully formulate the voter payoff structures that rationalize these two approaches to abstention in the \hyperref[sec:append]{Appendix}.

In both probabilistic and deterministic models of voting, the magnitude of a voter's net utility from the candidates, $c_1$ and $c_2$, is captured by the functional form of $u(.)$, especially when both candidates are positioned sufficiently close or far from the voter's ideal position. Formally, controlling for the spatial distance between the two candidates' positions, $|\delta_{i,c_1}-\delta_{i,c_2}|$, different 
functional form assumptions generate divergent predictions over the voter's indifference to the candidates as the voter's spatial distance to both candidates, represented by $\min\{\delta_{i,c_1},\delta_{i,c_2}\}$, changes. The predictions of the different functional form assumptions regarding how voter indifference between candidate choices changes as a function of $\min\{\delta_{i,c_1},\delta_{i,c_2}\}$ are as follows:
\begin{itemize}
\item \textbf{Linear Loss Function:} Voter indifference between candidates is not affected by $\min\{\delta_{i,c_1},\delta_{i,c_2}\}$.
\item    \textbf{Concave Loss Function:} Voter indifference between candidates decreases in $\min\{\delta_{i,c_1},\delta_{i,c_2}\}$.
\item    \textbf{Convex Loss Function:} Voter indifference between candidates increases in $\min\{\delta_{i,c_1},\delta_{i,c_2}\}$.
\item     \textbf{Reverse S-shaped Loss Function:} Voter indifference between candidates first decreases and then increases as $\min\{\delta_{i,c_1},\delta_{i,c_2}\}$ increases.
\end{itemize}

To uncover the functional form that best describes voter utility and hence observed voter behavior, we first identify empirical settings where the functional form assumptions generate contradicting predictions. These are electoral contexts where variation in spatial distance to both candidates can be observed across voter groups within the same race and for the same voter group across races. With data from the $2020$ US election, we identify two primary `cases' of these election types, which we refer to as Case $1$ and Case $2$. These cases and their divergent predictions are summarized in Table \ref{table_cases_des}. 
\begin{center}
\begin{table}
\caption{Functional Form Properties and Predictions}
\resizebox{\textwidth}{!}{ 
\fontsize{16}{20}\selectfont
     \begin{tabular}{|c|c|c|c|c|} 
        \hline 
        \textbf{\Large Functional Form} & \textbf{\Large Linear}&\textbf{\Large Concave}&\textbf{\Large Convex}&\textbf{\Large Reverse Sigmoid}\\ 
     \hline 
        \large{$\dfrac{d u(\delta_{i,c})}{{d \delta_{i,c}}}$}&
        \large{$<0$}&
        \large{$<0$}&
        \large{$<0$}&
        \large{$<0$}\\ 
        \hline
        \large{$\dfrac{d^2 u(\delta_{i,c})}{{d \delta_{i,c}}}$}&
        \large{$=0$}&
        \large{$<0$}&
        \large{$>0$}&
        \large{$<0$ if ${\delta_{i,c}}<t$ and $>0$ if ${ \delta_{i,c}}>t$}\\
        \hline
        \raisebox{20mm}{\textbf{\large Representation}} &\tikzpicone&\tikzpicsecond&\tikzpicthird&\tikzpicfourth\\
        \hline
\end{tabular}}
\label{table_func_forms}
\end{table}
\end{center}

In what follows, we provide a heuristic overview of Case $1$ and $2$. In this overview, for consistency with our empirical cases, we discretize the the uni-dimensional policy space into $n+1$ discrete segments such that the Democratic Party candidates take positions in segment $x_{c_D}\in0$, the Republican Party candidates take positions in segment $x_{c_R}\in n$ and the representative voter can be positioned at any segment $x_i\in\{0,1,2,...,n\}$.    

\begin{center}
\begin{table}    
\caption{\normalsize Case Setup and Predictions}
\fontsize{10}{14}\selectfont
\centering
\begin{tabular}{|c|c|c|} \hline 
\multicolumn{2}{|c|}{\textbf{\normalsize Set-up}} & 
\textbf{\normalsize Diverging Predictions} \\ \hline  
\textbf{\normalsize Case 1} &  
\makecell{Voting in races between \\ same party candidates} & 
\makecell{Changes in voter's indifference to the candidates \\ as the voter gets spatially distanced \\ from both candidates}\\ \hline  
\textbf{\normalsize Case 2} &  
\makecell{Voting in races between \\ opposing party candidates \\ with varying polarization levels} & 
\makecell{Changes in voter's indifference to the candidates \\ as the voter gets spatially distanced \\ from both candidates}\\ \hline 
\end{tabular}
\label{table_cases_des}
\end{table}

\end{center}

\subsection*{\textbf{Case 1}: Voting in races between same party candidates}

\begin{figure}[h!]
\begin{center}
\begin{tikzpicture}
    \coordinate (xCD1) at (0,0);
   \coordinate (xCD2) at (2,0);
    \coordinate (xi) at (4,0);
    \coordinate (xip) at (6,0);
    
 \draw (xCD1) -- (xip);
  \fill (xCD1) circle (2pt) node[below] {$x_{c_{D1}}$};
    \fill (xCD2) circle (2pt) node[below] {$x_{c_{D2}}$};
    \fill (xi) circle (2pt) node[below] {$x_i$};
   \fill (xip) circle (2pt) node[below] {$x_i'$};
   
    \draw[->, thick] (4,0.2) arc (180:0:1) node[midway, above] {$\Delta{x_i}$};
\end{tikzpicture}
\end{center}
\caption{Case 1 Position Shifts of $\Delta x_i$}
\label{fig_uni_dim_deltaC2}
\end{figure}
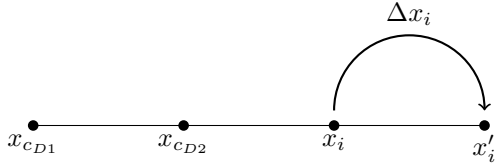

In races between candidates from the same party, i.e. two Democrats or two Republicans compete against each other, a change in the voter's ideal policy position, $x_i$, moves the voter closer or away from both candidates. Consider a race between two Democratic candidates, denoted by $c_{D1}$ and $c_{D2}$, occupying potentially distinct positions within the segment $x_{{c_{D1},c_{D2}}}=0$. Then, as voter $i$'s ideal position, $x_{i}$, shifts in the positive direction, $\min\{\delta_{i,c_{D1}},\delta_{i,c_{D2}}\}$ increases. An illustration of a representative voter $i$'s position shift in the uni-dimensional settings is presented in Figure \ref{fig_uni_dim_deltaC2}.

While the functional form assumptions share the prediction that voter utility from both candidates decreases as the spatial distance to both increases, their predictions diverge over the magnitude of net voter utility from, and indifference to, the candidates. Then, the parameter that reflects the divergent predictions from the different functional form assumptions is voter indifference, captured by the term $-|u(\delta_{i,c_{D1}})-u(\delta_{i,c_{D2}})|$, as a function of $\min\{\delta_{i,c_{D1}},\delta_{i,c_{D2}}\}$, or $x_{i}$. These divergent predictions are immune to the assumed probabilistic voting functions. Table \ref{func_form_properties_C2} summarizes the different predictions regarding voter indifference for each utility function. 

\begin{center}
\begin{table}
\caption{ Functional Form Properties and Predictions}
\resizebox{\textwidth}{!}{ 
\fontsize{10}{12}\selectfont
     \begin{tabular}{|c|c|c|l|} 
        \hline 
        \raisebox{-3mm}{\textbf{\normalsize Predictions over}:}&
        \raisebox{-3mm}{\textbf{\normalsize Predictions over}:}& 
        \raisebox{-9mm}{{ \normalsize \makecell{\textbf{Functional}  \textbf{Form}  }}}& 
        \raisebox{-9mm}{{ \normalsize \makecell{\textbf{Empirical}  \textbf{Predictions}  }}}\\
        \raisebox{9mm}{$\mathbf{-|u(\delta_{i,c_{D1}})-u(\delta_{i,c_{D2}})|}$} & 
        \raisebox{7mm}{$\mathbf{\dfrac{-|u(\delta_{i,c_{D1}})-u(\delta_{i,c_{D2}})|}{\Delta x_i}}$}& 
        &\\
        \hline 
         \tikzindiffoverKone &
        \raisebox{14mm}{$0$} &
        \raisebox{14mm}{Linear} &\raisebox{14mm}{\makecell{Voter indifference \\ stays the same}}\\ 
        \hline 
        \tikzindiffoverKtwo &
        \raisebox{14mm}{$>0$} &
        \raisebox{14mm}{Strictly Concave} &\raisebox{14mm}{\makecell{Voters becomes less \\ indifferent}}\\ 
        \hline
        \tikzindiffoverKthree &
        \raisebox{14mm}{$<0$} &
        \raisebox{14mm}{Strictly Convex} &\raisebox{14mm}{\makecell{Voter becomes more \\ indifferent}}\\
        \hline
        \tikzindiffoverKfour &
        \raisebox{14mm}{$>0$ and $<0$} &
        \raisebox{14mm}{Reverse S-shaped} &\raisebox{14mm}{\makecell{Voter becomes less \\indifferent below a \\minimum distance\\ threshold, and more \\ indifferent above \\ the threshold}}\\
        \hline
    \end{tabular}}
\label{func_form_properties_C2}
\end{table}

\end{center}

The linear loss function predicts that voter indifference to the candidates will be unaffected by position shifts that move the voter away from both candidates. Convex loss functions, on the other hand, predict voter indifference to be monotone increasing, while the concave loss functions predict voter indifference to be monotone decreasing. Reverse S-shaped loss functions predict voter indifference to first decrease and then increase as these differences become sufficiently large.  Depending on the exact policy positions of the two same party candidates, the convex and reserve S-shaped functional form can both explain increasing levels of indifference. But it is also possible to distinguish the predictions of the convex and reverse S-shaped loss functions in a set-upf where there exist voters who are sufficiently close to the position taken by both candidates which will be observed in \emph{Case 2}.

\subsection*{Case 2: Voting in races between opposing party candidates with varying polarization levels}

\begin{figure}[h!]
\begin{center}
\begin{tikzpicture}
    \coordinate (xCD) at (0,0);
    \coordinate (xCDp) at (2,0);
    \coordinate (xi) at (3.3,0);
    \coordinate (xCR) at (4,0);
    \coordinate (xCRp) at (6,0);

    \draw (xCD) -- (xCRp);

    \fill (xCD) circle (2pt) node[below] {$x_{c_{D'}}$};
    
    \fill (xCDp) circle (2pt) node[below] {$x_{c_{D}}$};
    \fill (xi) circle (2pt) node[below] {$x_i$};
    \fill (xCR) circle (2pt) node[below] {$x_{c_{R}}$};
    \fill (xCRp) circle (2pt) node[below] {$x_{c_{R'}}$};

    \draw[->, thick] (4,0.2) arc (180:0:1) node[midway, above] {$\Delta_{pol}$};
    \draw[->, thick] (2,0.2) arc (-180:0:-1) node[midway, above] {$\Delta_{pol}$};
\end{tikzpicture}


    
\end{center}
\caption{Case 2 Position Shifts of $\Delta pol$}
\label{fig_uni_dim_deltaC3}
\end{figure}
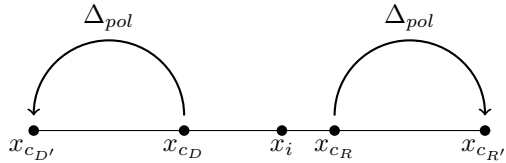

In \emph{Case 2}, we consider races between opposing party candidates, $c_D$ and $c_R$. But instead of variation in the ideal points of voters, we focus on a specific set of voters - voters who exhibit the largest level of indifference to the opposing party candidates across office races. We refer to this set of individuals as the `moderate' voter group. Then, we examine how indifference between the two opposing party candidates amongst moderate voters changes as polarization between the candidates increases. This case can be conceptualized as candidates $c_D$ and $c_R$ taking positions at $x_{c_D}\in0$ and $x_{c_R} \in n$, and the voter is positioned at the segment $x_i=\frac{n}{2}$. Here, instead of variation in the voter's ideal positions, what leads to variation in the term $\min\{\delta_{i,c_{D1}},\delta_{i,c_{D2}}\}$ is the polarization level of the candidates across races. Hence, we investigate the impact of candidate polarization, which is denoted by ${pol}$ and reflects an increasing distance to both candidates, ${pol} \approx \min\{\delta_{i,c_{D1}},\delta_{i,c_{D2}}\}$, on the level of indifference moderate voters exhibit between the two candidates. We provide an illustration of this kind of increasing candidate polarization for a specific group of voters in Figure \ref{fig_uni_dim_deltaC3}. Then, the predictions of the different functional form assumptions regarding moderate voter indifference to the candidates follow those presented in Table \ref{func_form_properties_C2}. 

As in Case $1$, while the reverse S-shaped loss function allows for non-monotone predictions, each of the other functional forms are monotone regarding voter indifference, and indifference levels among moderate voters either does not change, always increases or always decreases with increasing polarization. As candidate polarization increases, moderate voters are expected to first become less indifferent to the opposing party candidates and then become more indifferent, reaching their maximal indifference when polarization is at its highest point.

\section*{Data Description}

In this section, we provide more detailed background on our data and the respective measures upon which our analysis relies. In the U.S., county officials convert each ballot cast in an election into an official voting record by inserting the ballot into a scanner and tabulator machine. As a by-product of the tallying process, these machines provide electronic records of every choice made on every ballot cast in a U.S. election \citep{wack2019cast}. Election officials in turn often use these `cast vote records' (CVR) to conduct post-election audits. The anonymous CVR data has rarely been made publicly available to promote the transparency of their administration  \citep{kuriwaki2023still} and has been used in the literature studying voting behavior \citep{Gerber2004, Herron2007, kuriwaki_2023, kuriwaki23}.

After the 2020 election, a group of election skeptics, data scientists, and other interested parties engaged in a large-scale, crowd-sourced effort to collect CVR records from county offices, often by flooding them with FOIA requests \citep{green2023foia}. Jeffrey O'Donnell created a website to house a large collection of files containing these records, portions of which he used in a report on election fraud.\footnote{Various media outlets described his and others efforts to create this public database. See \emph{The Washington Post}, ``Trump backers flood election offices with requests as 2022 vote nears,'' September 11, 2022. \url{https://perma.cc/Z52R-KZB5}.}  Together with a group of Harvard and MIT researchers, we have worked cautiously to convert these electronic records into clean, validated, tidy data. See \citet{kuirwaki_cvr_dat} for more details on the steps we took to process and clean the data.\footnote{Summarizing the cleaning procedure briefly: we first processed each county-level file through multiple standardization procedures to create a single database of ballots. Note that this effort is non-trivial: CVRs vary greatly by the manufacturer of the voting tabulator, which in the U.S. are mostly Dominion, ES\&S,  and Hart. We have not identified any obvious data issues beyond records that are likely missing for administrative and privacy reasons. One small caveat is that in roughly ten percent of cases in our \emph{country-wide} sample, the CVR data split a voter’s record across multiple observations (i.e. separate rows in the data for different pages of the physical, paper ballot), and we are unable to reliably link voters across these split records.  Therefore, we study various subsets of this data given the availability of offices on the ballot. We show in \citet{kuirwaki_cvr_dat}  that the data remain representative.}

The CVR data provides a complete record of each voter's choices across the entire ballot, including how they cast their vote in the federal, state and local office races, as well as in their state's ballot measures. This enables us to connect the individual voters' policy positions with respect to the Democratic and Republican Parties and their candidate and party choices. We will be using the state ballot measure which are typically endorsed by one of the two major parties while opposed by the other, to establish an ordinal measure of how voters are positioned relative to the two major parties in their state. We call this measure \emph{Voter Group}, and describe how it is defined in the following two sections. Then, we investigate which functional form best explains the voters' ballot choices in office races -including which if any races they choose to abstain from casting a vote in-  as a function of their proximity to each major party's policy position in the three \emph{Cases} we outlined. 

As shown in Table \ref{ca_cover}, the total number of  voters in our California sample is $14.8$ million for the federal offices, and $7.1$ million for the state ballot propositions we can link to at least one of these federal offices. Note that this does not indicate that $50$ percent of voters in California abstained from the proposition votes - a large portion of the difference is due to ballot fragmentation, as we do not retain observations where we cannot connect at least one office vote to proposition vote.  Of those $7.1$ million, roughly $6$ million, or $84$ percent, voted on all twelve of the ballot propositions - i.e. did not abstain from any proposition vote. In Colorado almost exactly the same number, $82$ percent, of voters voted on all eleven of the ballot propositions, which is shown in Table \ref{co_cover}.  In sum, we end up with $6$ million complete ballot records for California, when we account for fragmented ballots and discard voters who abstained from at least one proposition, and  $2.6$ million for Colorado.

\begin{table}[H]
\centering
\caption{Statewide Ballot Measures}
\scriptsize
\label{prop_description} 
\begin{tabular}{l l ll r}
\toprule
Measure & Description  &Source& Party Support & Result \\
\midrule
\multicolumn{4}{c}{\it California} & \\ [.05in]
Prop 14 & Stem cell research bonds  &Signatures& D & 51.1 \\
Prop 15 & Increase commercial and industrial property tax rates  &Signatures& D & 48.0 \\
Prop 16 & Expand affirmative action (repeal Prop. 209)  &Legislature& D & 42.8 \\
Prop 17 & Restore voting rights to felons on parole  &Legislature& D & 58.6 \\
Prop 18 & 17 year-olds may vote in primary if 18 at time of genl  &Legislature&  D & 44.0 \\
Prop 19 & Residential property tax rules for seniors \& inheritors  &Legislature& N & 51.1 \\
Prop 20 & Parole restrictions  &Signatures& R & 38.3 \\
Prop 21 & Expand rent control  &Signatures& D & 40.1 \\
Prop 22 & Ride-share drivers are independent contractors  &Signatures& R & 58.6 \\
Prop 23 & Dialysis clinic rules  &Signatures& N & 36.6 \\
Prop 24 & Expand consumer protections  &Signatures& D & 56.3 \\
Prop 25 & Bail reform   &Signatures& D & 43.6 \\
\midrule
\multicolumn{4}{c}{\it Colorado} & \\ [.05in]
Amend 76 & State that only `citizens' may vote  && R & 62.9 \\
Amend 77 & Allow certain cities to allow gambling  && N & 60.5 \\
Amend B & Repeal gallagher amend on property tax assessments  && D & 57.5 \\   
Amend C & Rules for gambling in charitable organizations  && N & 52.3 \\
Prop 113 & Enter state into natl popular vote pact for pres  && D & 52.3 \\ 
Prop 114 & Reintroduce gray wolves on public lands  && N & 50.9 \\ 
Prop 115 & Ban abortion after 22 weeks  && R & 41.0 \\ 
Prop 116 & Decrease income tax rate by .08 percent  && R & 57.9 \\ 
Prop 117 & Exempting new large enterprises from TABOR  && R & 52.5 \\ 
Prop 118 & Establish program for paid medical \& family leave  && D & 57.7 \\ 
Prop EE  & Increase tax on tobacco, earmark for health \& education  && D & 67.6 \\
\bottomrule
\end{tabular}
\end{table}

\newpage

\begin{table}
\centering
\caption{California CVR Data Coverage}
\footnotesize
\label{ca_cover} 

\begin{tabular}{rr}
\toprule
\multicolumn{1}{l}{\textbf{California 2020 Voter Population}} & 
 17,512,265\\ \\\midrule
\multicolumn{1}{l}{\textbf{CVR Sample}} & \\
Ballots for President  & 14,810,952
\\
 \mgray{of those}, linked records for 12 ballot measures & {7,172,442
}\\
 \mgray{of those}, no ballot measure abstentions & {6,050,053
}\\
 \mgray{of those}, linked to $\geq$ 3 contested partisan offices & 4,044,524 \\
\bottomrule
\end{tabular}
\end{table}

\begin{table}
\centering
\caption{Colorado CVR Data Coverage}
\footnotesize
\label{co_cover} 
\begin{tabular}{rr}
\toprule
\multicolumn{1}{l}{\textbf{Colorado 2020 Voter Population}} & 
 3,291,661\\ \\\midrule
\multicolumn{1}{l}{\textbf{CVR Sample}} & \\
Ballots for President  & 3,277,791
\\
 \mgray{of those},  linked records for 12 ballot measures & {3,173,136
}\\
 \mgray{of those}, no ballot measure abstentions & {2,623,235
}\\
 \mgray{of those}, linked to $\geq$ 3 contested partisan offices & 
 2,620,537
\\
\bottomrule
\end{tabular}
\end{table}

\subsection*{State Ballot Measures}
Ballot measures are a common feature in U.S. elections at both the state and local level.\footnote{For example, all states except Delaware, require voter approval of state constitutional amendments.} Although most states have included a state-level measure on their ballot at some point in their electoral history, the number of ballot measures in any given election varies widely. In two states, Colorado and California, a sizable number of statewide measures made it on the ballot in the 2020 general election. In 2020, Colorado residents voted on eleven and California voters on twelve ballot measures.

Like many states, ballot measures in California and Colorado come in at least two forms - legislative measures and initiative measures. The former are placed on the ballot by the state legislature if they can reach a majority vote. Initiative measures, on the other hand, can be placed on the ballot by non-elected officials through the ``initiative process". After a measure is written, with enough signatures from public the proposition appears on the ballot.\footnote{The number of signatures required changes every gubernatorial election, based on the percentage of how many votes were cast for governor, and by type of initiative - initiative constitutional amendments require about one third more than initiative statues and referendums. } 

In 2020, the statewide ballot measures on the California and Colorado state ballot together represent a wide range of social and economic issues, including whether to raise property taxes, expand rent control, ban cash bail, further protect consumer data privacy, resurrect affirmative action, change constitutional voting rights and restrict abortion rights. Various groups, both partisan and nonpartisan, provide endorsements for specific measures. The Democratic Party endorsed eight of the twelve statewide ballot propositions in California, while the Republican Party endorsed two.\footnote{Note that we exclude Proposition 19 and Proposition 23 from our analyses, as these propositions received conflicting endorsements from different divisions of the California Democratic Party. For example, the Santa Monica Democratic Club formally opposed Proposition 23 while the California Democratic Party endorsed it.} In Colorado, the two major parties took positions on nine of the eleven propositions.\footnote{ In Colorado, two propositions on gambling (Amendment 77 and Amendment C) were not given clear endorsements from either party, so we exclude them from our analysis.} The Republican party proposed or endorsed four and the Democratic Party proposed or endorsed five. Both parties also officially opposed the propositions endorsed by the other party. 
 Table \ref{prop_description} provides a complete list with detail on the topic, official outcome, and party support for each ballot proposition in the two states.

\subsection*{\textit{Empirical Measures}}
\label{sec:measure}

We utilize the $2020$ state ballot measures in California and Colorado to construct an ordinal measure of voters' preferred policy positions with respect to the positions taken by the Democratic and Republican Party.  In this exercise, instead of attempting to scale voters' ideal positions cardinally, we aim to group voters such that the average voter in each voter group can be ordinally ranked based on their proximity to the two major parties. Since our measure is constructed with state-level ballot measures, it must be noted that the voter group ranks in the two states are not comparable. 

\begin{figure}[h!]
\centering
\begin{minipage}{0.47\textwidth}
    \centering
    \includegraphics[width=\textwidth]{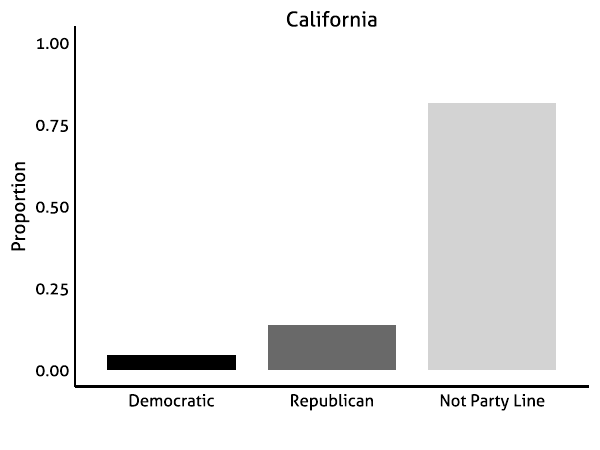}
\end{minipage}
\begin{minipage}{0.47\textwidth}
    \centering
    \includegraphics[width=\textwidth]{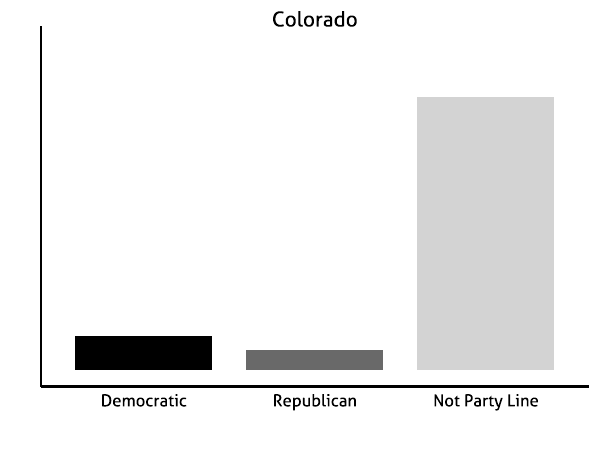}
\end{minipage}
\vspace{-0.5cm}
\caption{Proportion of Voters by Voting Pattern on Statewide Ballot Propositions}
\label{fig_partline}
\end{figure}

Using the votes cast on each ballot measure, each voter $i$ can be assigned a position vector $\vec x_{i}=(d_1, d_2,,...,d_n)$ in the policy space where $d_n$ represents the position taken on proposition $n$. As in both states the Democratic and Republican Parties took opposing positions in all ballot propositions, we denote the vote cast on each proposition $n$ as $d_n=\{0,1\}$ such that a vote that aligns with the Republican Party position corresponds to $d_n=1$ and a vote that aligns with the Democratic Party position corresponds to $d_n=0$. Then, the voters who align most with the Democratic Party share a position vectors $(0,0,0,..,0)$ and those align most with the Republican Party share a position vectors $(1,1,1,..,1)$.\footnote{Numerous interest groups took positions on the ballot measures and their positions generally match those of the parties. Using the positions taken by the interest groups in addition to the parties does not change our measure.} Importantly, relatively few voters vote the straight party line on the ballot propositions in both California and Colorado, as shown in Figure \ref{fig_partline}.

In our pursuit to infer voters' relative distance to the party positions, we construct a grouping measure that we call \emph{Voter Group} and denote their rank by $k$. We group voters into \emph{Voter Group}s based on the number of times they agreed with the Republican Party position, or equivalently disagreed with the Democratic Party position, across the ballot measures. Within each state, the rank of a \emph{Voter Group} is given by $k=\sum_n d_n$ such that the \emph{Voter Group} that most aligns with the Democratic Party position is  $k=0$ and the \emph{Voter Group} that most aligns with the Republican Party position is  $k=n$ where $n=10$ in California and $n=9$ in Colorado. It is important to emphasize that we do not claim that \emph{Voter Group}s to be composed of voter with identical policy preferences. But rather, that the average voter in \emph{Voter Group} $k+1$ is spatially positioned closer to the Republican Party and further away from the Democratic Party compared to the average voter in \emph{Voter Group} $k$.  

\begin{figure}[h!]
\centering
\begin{minipage}{0.49\textwidth}
    \centering
    \includegraphics[width=\textwidth]{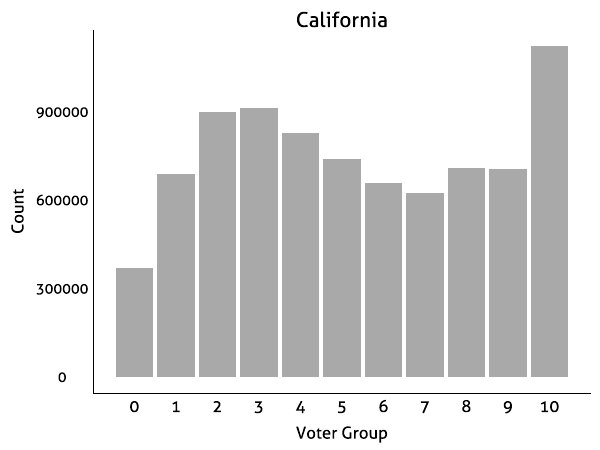}
\end{minipage}
\begin{minipage}{0.49\textwidth}
    \centering
    \includegraphics[width=\textwidth]{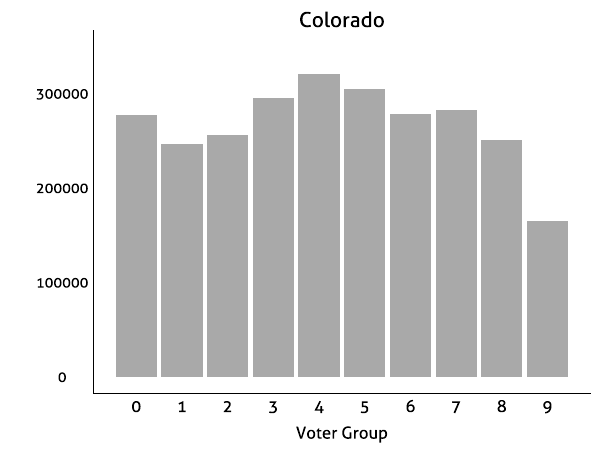}
\end{minipage}
\vspace{-0.1cm}
\caption{Distribution of Voter Group by State}
\label{fig_pos_dist_bar}
\end{figure}

It is also important to note that being in \emph{Voter Group}s $0$ and $n$ does not indicate that voters in these \emph{Voter Groups} perfectly align with a party position but only that these are the voters that most align with the party positions. For example, as the size of California's \emph{Voter Group} $k=10$ suggests, in Figure \ref{fig_pos_dist_bar}, the ballot measures do not capture as much variation in conservative, Republican leaning, positions. This might be due to the fact that only two of the ten ballot measures were proposed by conservative groups and endorsed by the Republican Party while eight are proposed by the liberal groups and the Democratic Party in California.

Though simple and intuitive, our group measure is highly correlated with the more complex scaling methods that aim to generate cardinal scales of preferred voter positions \citep{heckman_snyder}. These include optimal classification \citep{poole2000nonpar}, \citep{poole2005spatial}, \citep{rosenthal2004analyzing}, principle component analysis (PCA), and item response theory (IRT) models which are also highly correlated among themselves despite relying on substantively different assumptions over voter preference, e.g. functional form of voter utilities. 

Recall that there are $1024$ unique position vectors in California and $512$ in Colorado and the \emph{Voter Group}s are formed by amalgamation of voters with distinct position vectors. For example, two substantively different \emph{Voter Subgroup}s in California that amass voters with position vectors $(0 0 0 0 0 1 1 1 1 1)$ and $(1 1 1 1 1 0 0 0 0 0)$ are both placed in the same \emph{Voter Group},  $k = 5$. We will refer to the group of voters who share the exact position vector as \emph{Voter Subgroup}s and denote each by $sg\in k$, such that each \emph{Voter Group} $k$ is composed of ${n \choose k}$ \emph{Voter Subgroup}s. While we do not rank them, we denote the \emph{Voter Subgroup} pair that only differ in their response to a specific proposition $n$ by $sg^+_n$ and $sg^-_n$ such that the former group took a position $d_n=1$ and the latter $d_n=0$ on measure $n$, and they align in their responses to all other ballot measures. We will rely on this \emph{Voter Subgroup} measure to check the robustness of our results to the heterogeneities at the \emph{Voter Group}s.

We denote each office race by $j$ and the type of race each race $j$ belongs to, which can be presidential, congressional, legislative, etc, is denoted by $J$ such that $j\in J$. The number of voters in \emph{Voter Group} $k$  is given by $N_{k}$ and the number of voters in \emph{Voter Group} $k$ who vote in race $j$ is $N^{j}_{k}$. Then, the total number of voters in \emph{Voter Group} $k$ who voted for a the Democrat Party, Republican Party or abstained in race $j$ will denoted by $N_{k}^{j}(D)$, $N_{k}^{j}(R)$, $N_{k}^{j}(A)$, respectively.

To operationalize voter indifference, which we formally represent as $-|u(\delta_{i,c_1})-u(\delta_{i,c_2})|$, we employ two common measures: abstention and vote predictability. First, the probability that a randomly picked voter in voter group $k$ abstains in a race $j$ is captured by the average abstention rates, $Pr_{k}^j(A)=\frac{N_{k}^{j}(A)}{N_k^j}$. Then, the overall probability that a voter from voter group $k$ abstains is take by $\frac{\sum_j Pr_{k}^j(A)}{\sum_j 1_j}$ where $1_j$ is the indicator for each race.

We rely on the across-race averages instead of weighting race averages by the number of votes cast on that race within \emph{Voter Group}s. We acknowledge that the probability of voting for a particular party in a particular race is heterogeneous for individuals within each \emph{Voter Group}, or put differently, \emph{Voter Group} is not the only predictor of voter behavior. We draw on additional factors, such as how individuals vote in other in other races and race level shocks, to decompose our analysis of \emph{Voter Group} behavior across these other important predictors using race individual level fixed effects and find consistent results.

Our second indicator of indifference is vote predictability of a randomly picked voter from a \emph{Voter Group}. As the margin of support for the preferred candidate increases within a particular \emph{Voter Group}, we expect the average voter from this \emph{Voter Group} to be less indifferent to the opposing party candidates in race $j$. Our measures of vote predictability for \emph{Voter Group} k in race $j$ is $|\frac{N_{k}^{j}(D)}{N_k^j}-\frac{N_{k}^{j}(R)}{N_k^j}|$ where $\frac{N_{k}^{j}(D)}{N_k^j}$ is the average vote for the Democratic Party candidates within each \emph{Voter Group} $k$ corresponding to our formal measure $|Pr_i(D)-Pr_i(R)|$.\footnote{In races with more than one candidate, such as the US presidential race, we base the vote predictability measure on the vote margin between the candidates with the greatest cumulative vote share.} Then, the overall vote predictability among voters in \emph{Voter Group} $k$ is:

$$\frac{\sum_j|\frac{N_{k}^{j}(D)}{N_{k}^j}-\frac{N_{k}^{j}(R)}{N_{k}^j}|}{\sum_j 1_j}$$

\begin{figure}[h!]
\centering
\begin{minipage}{0.42\textwidth}
    \centering
    \includegraphics[width=\textwidth]{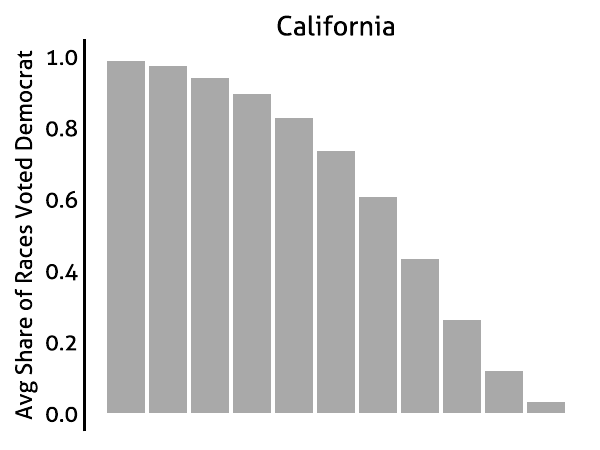}
\end{minipage}
\begin{minipage}{0.42\textwidth}
    \centering
    \includegraphics[width=\textwidth]{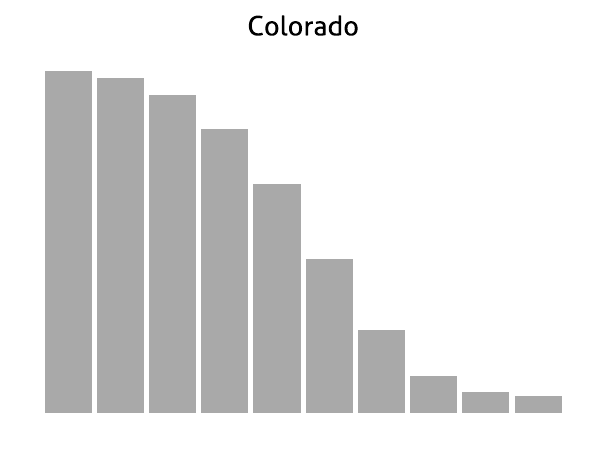}
\end{minipage}
\vspace{-0.2cm}
\caption{Average Share of Races Voted Democrat by \emph{Voter Group} for California (left) and Colorado (right)}
\label{figbar_VGroup_DemRepVShare}
\end{figure}
\begin{figure}[h!]
\begin{center}
 \includegraphics[scale=0.75]{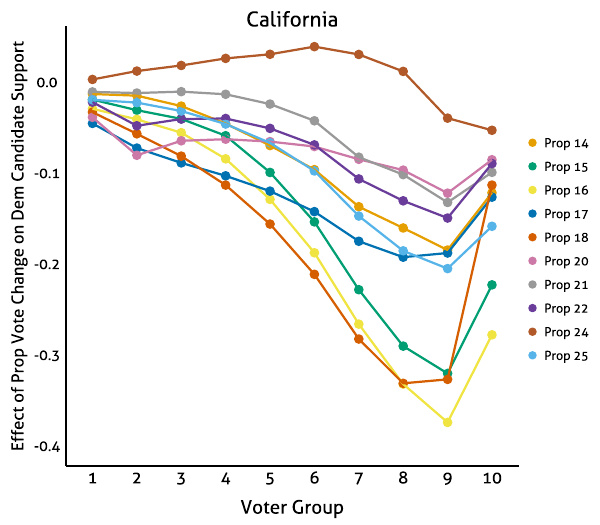}
 \includegraphics[scale=0.75]{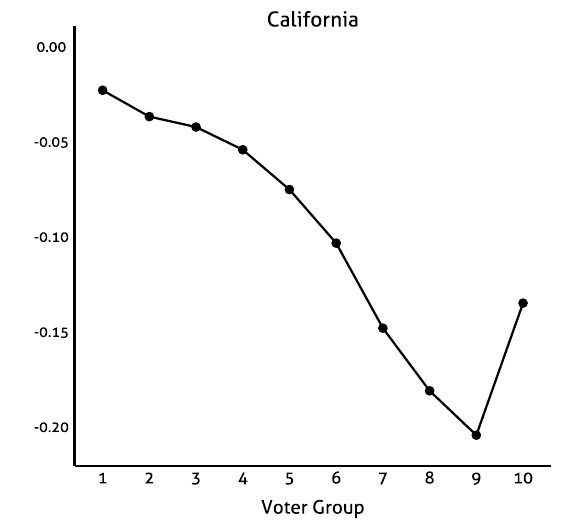}
  \includegraphics[scale=0.75]{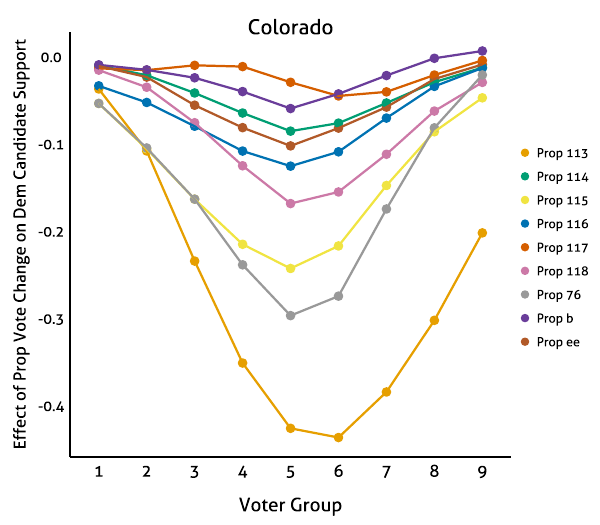}
  \includegraphics[scale=0.75]{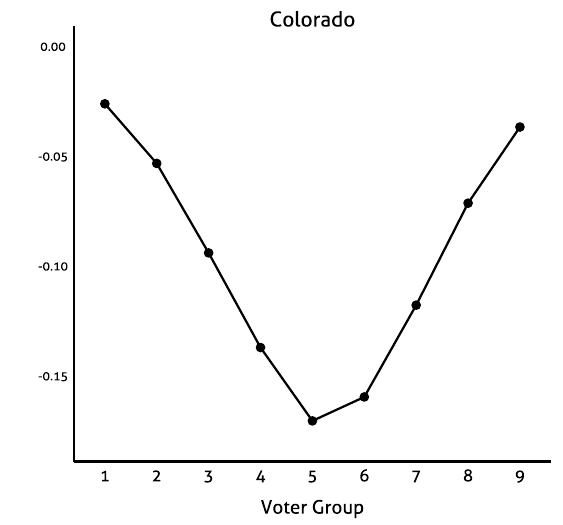}
\end{center}
\vspace{-0.3cm}
\caption{Average Effect of Change in Specific Proposition Responses Among Voter Groups on Support for Democratic Candidates in Democratic vs. Republican Races}
\label{fig_ballot_diffDR}
\end{figure}

It is important to note that while our simple \emph{Voter Group} measure explains roughly 50\% of the variation in candidate choice for races where both parties are on the ballot, the more nuanced \emph{Voter Subgroup} measure only increases this marginally, and in some races, not at all, which we show in Table \ref{fig_indif_wrace} of the \hyperref[sec:append]{Appendix}. Further, the monotone trend of decreasing support for the Democratic Party candidate shown in Figure \ref{figbar_VGroup_DemRepVShare} provides rudimentary support for our \emph{Voter Group} measure and the spatial models of voting. These trends hold not only at the \emph{Voter Group} level but also at the level of each individual ballot measure. Figure \ref{fig_ballot_diffDR}, we display the percent change in Democratic Party support at the \emph{Voter Subgroup} level as a result of a change in individual ballot measure(s) support. That is, we keep support constant across each of the ballot measures except one, and compare Democratic Party support between \emph{Voter Subgroup}s, in neighboring \emph{Voter Group}s, which only differ in their support for one of the ballot measures. 

\begin{figure}[h!]
\centering
\begin{minipage}{0.49\textwidth}
    \centering
    \includegraphics[width=\textwidth]{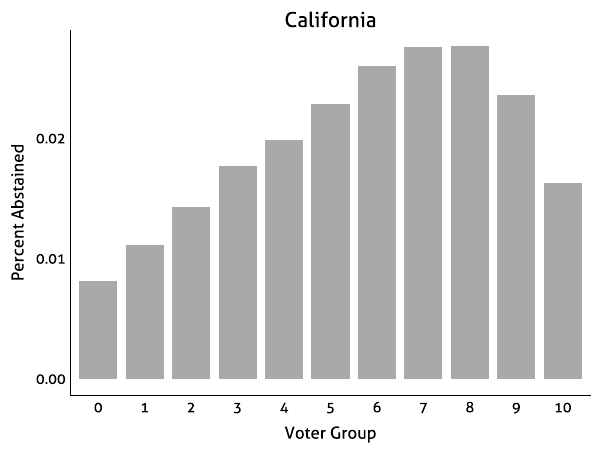}
\end{minipage}
\begin{minipage}{0.49\textwidth}
    \centering
    \includegraphics[width=\textwidth]{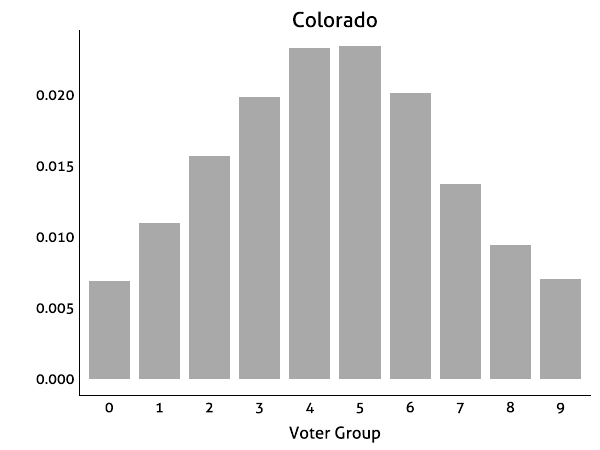}
\end{minipage}
\vspace{-0.1cm}
\caption{Distribution of Abstention in State and National Offices by State}
\label{fig_pos_dist_bar_roll}
\end{figure}

\begin{figure}[h!]
\centering
\begin{minipage}{0.49\textwidth}
    \centering
    \includegraphics[width=\textwidth]{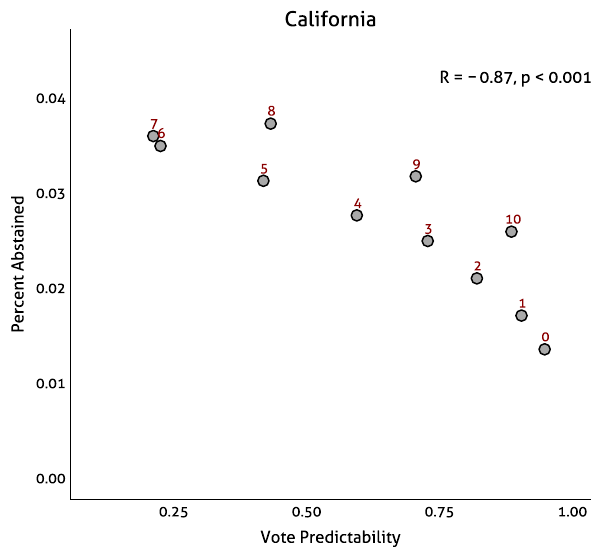}
\end{minipage}
\begin{minipage}{0.49\textwidth}
    \centering
    \includegraphics[width=\textwidth]{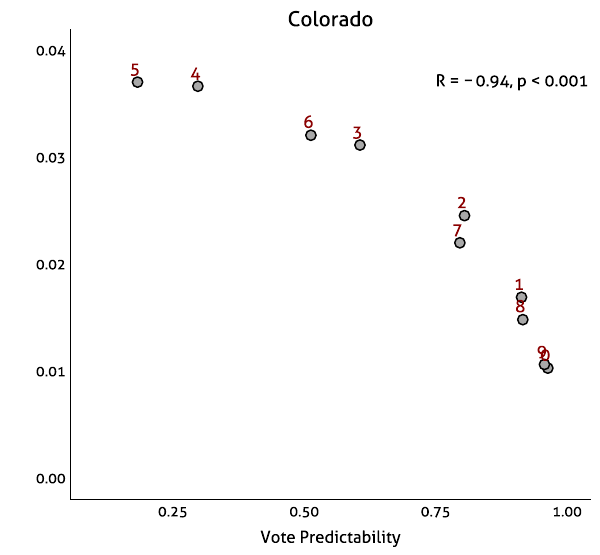}
\end{minipage}
\vspace{-0.1cm}
 \caption{Correlation Between Two Measures of Voter Indifference: Abstention and Vote Unpredictability in Republican vs. Democrat Races}
\label{fig_indiff_corr_ideoAvg_DR}
\end{figure}

Across races, the \emph{Voter Group}s most indifferent to the opposing party candidates are the groups that are positioned in the middle of, or at an equal distance to, the two opposing party positions. We show this is consistent for both measures of indifference: abstention behavior and vote predictability. Figure \ref{fig_pos_dist_bar_roll} displays percent of voters who abstained within each \emph{Voter Group} across all national and state level races where a Republican and a Democrat appeared on the ballot in California and Colorado, respectively.  We highlight a strong, negative correlation between the two measures of indifference, such that individual in \emph{Voter Group}s with relatively low vote predictability also tend to abstain at higher rates, as shown in Figure \ref{fig_indiff_corr_ideoAvg_DR}. We also demonstrate that the negative correlation between abstention rates and vote predictability holds within \emph{Voter Group}s across races in the \hyperref[sec:append]{Appendix}.\footnote{Figure \ref{fig_indiff_corr_rep_CO} shows the trend is very strong for all US House races, and equivalent figures can be found in the Appendix for the State House (Figure \ref{fig_indiff_corr_StHou_CO}) and the State Senate (Figure \ref{fig_indiff_corr_StSen_CO}). The correlation is also shown for both states in the Appendix in Table's \ref{corr_indiff_table_CO} and \ref{corr_indiff_table_CA}}. That is, in races where a specific \emph{Voter Group} exhibits high rates of abstention, we also observe an overall lower vote predictability of this group. This exercise serves the dual purpose of demonstrating the spatial models' and our \emph{Voter Group} measures' predictive power over voting behavior, as well as corroborating that our two measures of indifference are indeed capturing the same phenomenon.

\section*{Empirical Results}

We now proceed with a series of analyses regarding voter indifference in races with same party and opposing party candidates, respectively. In these two analytical cases, we accumulate evidence that voter loss follows the reverse S-shaped function.

\subsection*{Case 1: Voter Behavior in Races With Candidates From The Same Party}
\label{sec:empiric_case2}
Races between candidates from the same party present the opportunity to observe how voters behave when their spatial distances to both candidates changes in the same direction and test the diverging predictions of competing functional forms. In this setting, our choice of functional form is substantively consequential for how we understand voter indifference. Under the convex or reverse S-shaped functions, voters are predicted to become more indifferent to both candidates as their distance to both grows, while the concave function predicts the opposite: that voters in this setting experience decreasing levels of indifference. If we observe voters becoming more (less) indifferent, we can establish strong evidence for (against) the convex or reverse S-shaped functions, and against (for) the concave function. More specifically, we should only observe voters becoming more indifferent to both candidates if voter loss follows the convex or reverse-sigmoid functional form.\footnote{The convex and the reverse S-shaped functional forms predict higher indifference as voter groups' spatial distance to both candidates increases.} 
 
Our CVR data from the state of California allows us to examine voter behavior in this context because with the passage of Proposition 14 in 2010, the way primary elections work in California changed to the current ``top two" system. Unlike in the majority of US states, political parties are no longer responsible for nominating a candidate to run for state and Congressional offices in the general election.\footnote{In addition to California, exceptions to the closed primary system include the state of Washington, Louisiana (often referred to as a jungle primary), and as of 2020, Alaska.} Instead, all candidates are pooled in one primary. The two candidates with the most votes advance to the general election. In 2020, a total of seven US House, three State Senate and eleven State House races fielded two Democratic candidates in the general election.\footnote{The races where two Democrats appeared on the ballot in the 2020 general election  include California's 12th, 18th, 29th, 34th, 38th, 44th, and 53rd congressional Districts, State Senate District's 11, 15, and 33, and State House District’s 10, 13, 20, 46, 50, 53, 54, 59, 63, 64. Note that these same jurisdictions were redistricted in 2020 and now belong to different congressional districts.} Just two Republican vs Republican races took place in 2020 statewide, in State House district's 33 and 38.  We conduct the main analysis on all same party races but with a direct focus on the races between Democrats because their sheer quantity facilitates broader generalization.

In these races, we consider both candidates to have positions in our policy space close to the Democratic Party position, $k=0$. As the rank of \emph{Voter Group} increases, the spatial distance between the voters within the \emph{Voter Group} and both of the Democratic candidates increases. Reversely, in the few instances where we observe two Republican candidates on the ballot, as the rank of \emph{Voter Group} decreases, these voters' spatial distance to both Republican candidates increases. 

We begin this analysis by assessing indifference at the group and individual level through an examination of abstention behavior. Here we measure indifference at the voter-race level as a binary indicator that takes $1$ if the voter abstained in that particular race, and 0 otherwise. Taking abstention as an indicator of indifference, we observe that voters do indeed grow more indifferent to both candidates as their spatial distance to both candidates increases. Specifically, we show that at the \emph{Voter Group} level, average abstention increases with each additional shift away from the candidates' party position. 

\begin{figure}[h!]
\begin{center}
 \includegraphics[scale=0.80]{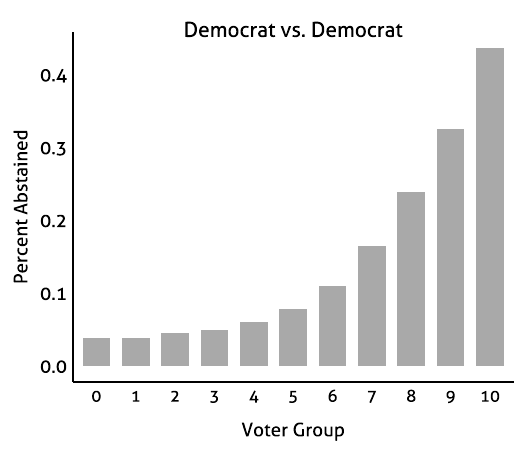}
 \includegraphics[scale=0.8]{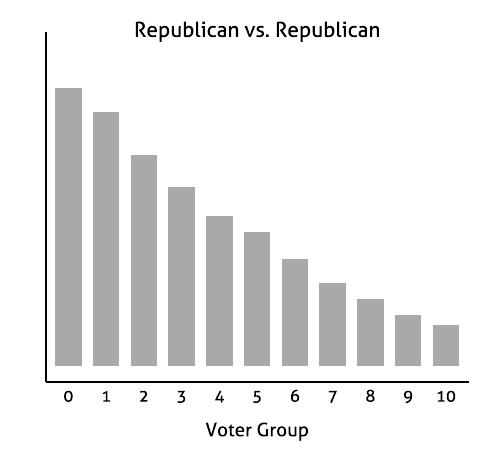}
\end{center}
 \vspace{-0.45cm}
 \caption{Abstention in Democrat v. Democrat races (left) and Republican v. Republican Races (right), California 2020}
 \vspace{-0.2cm}
\singlespacing
\small
 \textit{Note}: The barplots display the percent of individuals who abstained within each \emph{Voter Group}, where the figure on the left labelled `Democrat vs. Democrat' shows the average across all the races where two Democrats appeared on the ballot, while the right most figure, labelled 'Republican vs. Republican' shows the average across the State House races where a Republican faced a Republican for each value of \emph{Voter Group}.  In $2020$, two State House races - District's $33$ and $38$ - featured two Republican candidates.  See Figure \ref{fig_dd_roll_item} for the complete list of Democrat vs. Democrat races in $2020$.
  \label{fig_DD_RR_roll}
\end{figure}

\begin{figure}[h!]
 \includegraphics[scale=0.99]{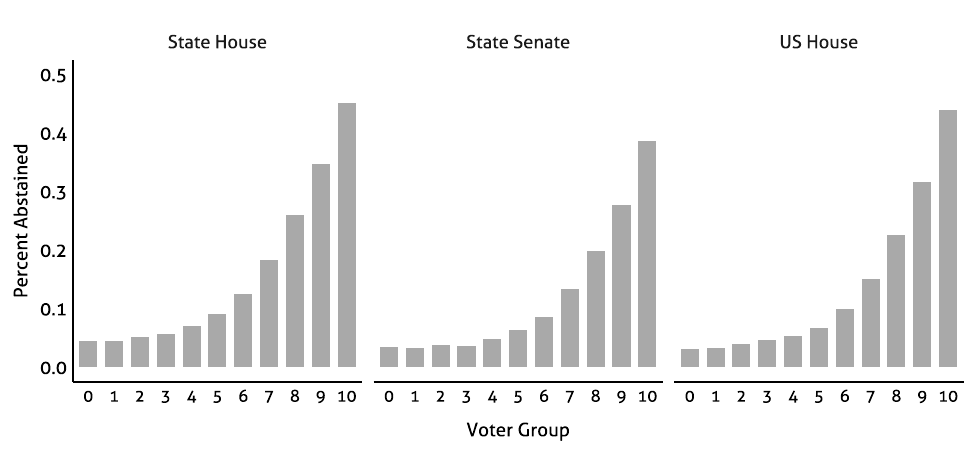}
\vspace{-0.45cm}
 \caption{\textbf{Abstention in Democrat v. Democrat Races by Office Type, California 2020}}
 \vspace{-0.35cm}
\singlespacing
\small
 \textit{Note}: The barplots display the percent of individuals who abstained in the House and State House races where a Democrat faced a Democrat for each value of \emph{Voter Group}. In $2020$, seven US House races - District’s $12, 18, 29, 34, 38, 44$, and $53$; eleven State House races - District’s $10, 13, 20, 46, 50, 53, 54, 59, 63, 64, 78$; and three State Senate races - District’s $11, 15$, and $33$, featured two Democratic candidates. Please see the Data Description section for complete details on the \emph{Voter Group} measure, including the ballot measure language, partisan support, and complete distribution.
 \label{fig_dd_roll_item}
\end{figure}

This relationship is strong and consistent across an impressive set of race-level features -  including office type, candidate party ID, and voter's inferred party ID. In Figure \ref{fig_DD_RR_roll}, we display the abstention trends across \emph{Voter Group}s in Democrat-Democrat races side-by-side with Republican-Republican races. Figure \ref{fig_DD_RR_roll} demonstrates that as the voters' spatial distance increases to both candidates, so does their indifference to the candidates.
Focusing on races between two Democratic Party candidates due to the larger number of races of this sort, we demonstrate that observed abstention trends across \emph{Voter Group}s hold consistently across different race types, the US House, the State Senate, and the State House, in Figure \ref{fig_dd_roll_item}. For completeness, we also show that this pattern holds for each individual Democrat vs. Democrat race in Figure \ref{fig_dd_roll_race} of the \hyperref[sec:append]{Appendix}, which reveals stark contrast to abstention behavior in races between a Democratic and a Republican Party candidates which is shown in Figure A.3 of the Appendix.

\begin{figure}[h!]
\begin{center}
 \includegraphics[scale=0.94]{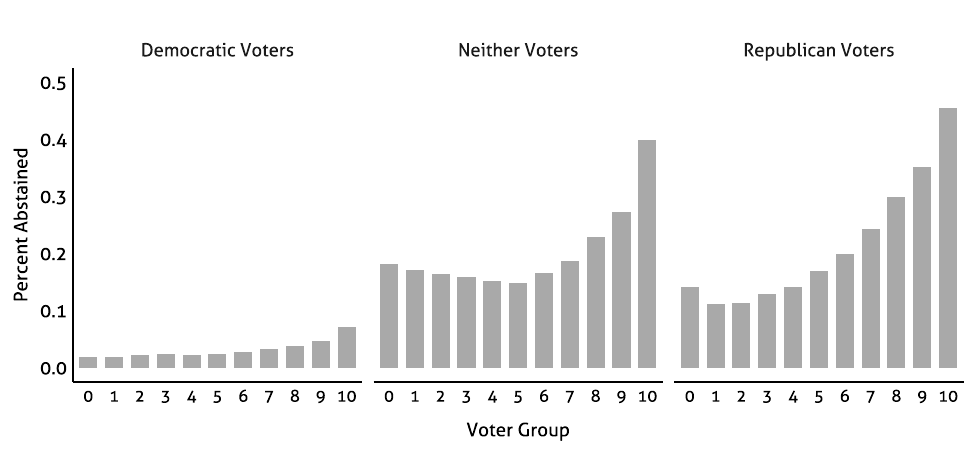}
\end{center}
 \vspace{-0.45cm}
 \caption{Abstention in Democrat v. Democrat Races by Voter Type}
 \vspace{-0.2cm}
\singlespacing
\small
 \textit{Note}: The barplots display the percent of individuals who abstained in the US House, State House, and State Senate races where a Democrat faced a Democrat for each value of \emph{Voter Group}. The left most and right most plots display individuals who voted Democrat and Republican in all other races where they could, respectively. The middle plot displays individuals who did not vote either Democrat or Republican in all other races where they could, and so a group of voters who split their ticket between Republican and Democratic candidates, and/or between one of the two major parties and a third party. 
  \label{fig_dd_roll_type}
\end{figure}

Further, we show that these trends do not arise merely due to the voters' party loyalties. That is, in races between two Democratic Party candidates, the abstention rates increase with the \emph{Voter Group} rank independent of which party these voters voted in other races. In Figure \ref{fig_dd_roll_type}, we pool all the Democrat vs Democrat races across office type and then create the same plot for three distinct voter types: Democratic Voters, Neither Voters, and Republican Voters. The groups of Democratic and Republican Voters are comprised of individuals who voted Democrat and Republican candidates in all other races,  respectively, where they were given the option to vote for either party candidate\footnote{To be clear, to create this measure of voter type, we subset to just the voters in our sample who were eligible to vote in at least three additional partisan offices where a Republican and a Democrat appeared on the same contest. That is, we made sure our measure of voter type was appropriately inclusive so only accounted for races where the voters COULD have feasibly voted Democrat or Republican, and was based on at least three observations where we could observe this. We were of course limited by the restricted number of partisan races in California, where offices below the state House and Senate are non-partisan.}. The Neither Voters group includes those voters who do not belong in the Democratic or Republican Voter categories, i.e. they did not vote all Democrat or all Republican in all other races where they could and voters who split their ticket between Republican and Democratic candidates. Figure \ref{fig_dd_roll_type} demonstrates that even those voters who give no indication of party loyalty grow more indifferent as their spatial distance to the candidates increases.

\begin{figure}[h!]
\begin{center}
 \includegraphics[scale=0.75]{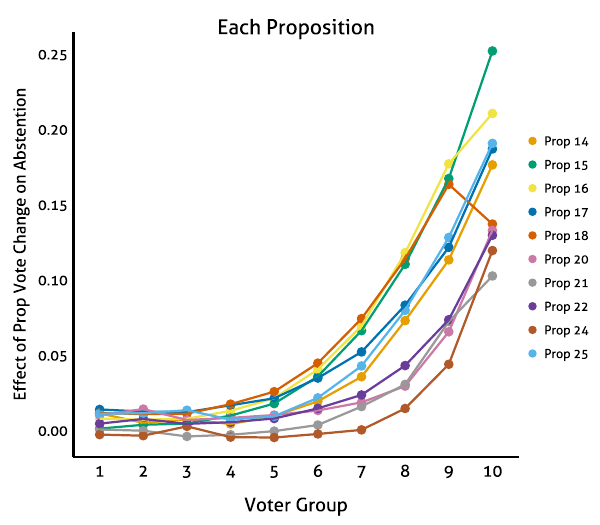}
  \includegraphics[scale=0.75]{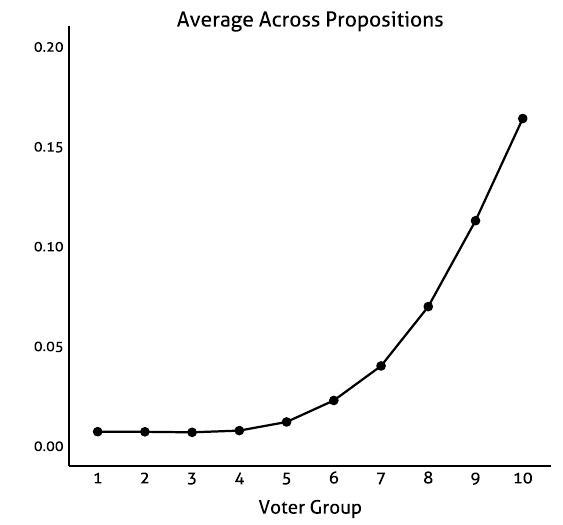}
\end{center}
 \vspace{-0.45cm}
\caption{Average Effect of Change in Specific Proposition Responses Amongst Voter Groups on Abstention in Democrat vs. Democrat Races}
\label{fig_ballot_regDD}
\end{figure}

\begin{figure}[h!]
\begin{center}
 \includegraphics[scale=0.9]{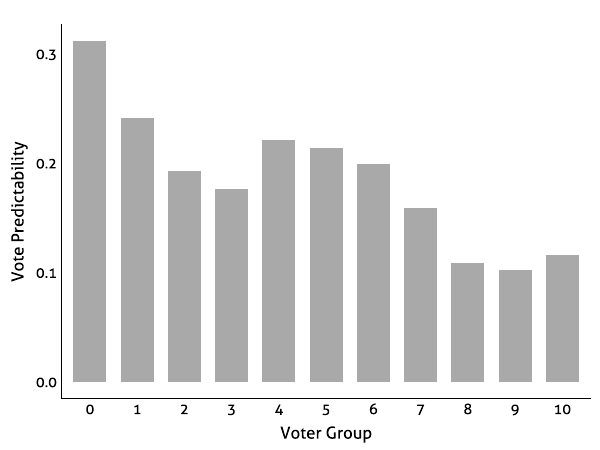}
\end{center}
 \vspace{-0.45cm}
 \caption{Distribution of Voter Predictability in Democrat vs. Democrat Races}
\label{fig_VPred_Overall}
\end{figure}

Importantly, we also need to assess whether these same patterns hold with our alternative measure of indifference: vote predictability. First, in Figure \ref{fig_VPred_Overall}, we also show distribution of vote predictability across \emph{Voter Group} in races between two Democratic Party candidates. While the decreasing vote predictability is only supported by the convex and reserve S-shaped functional forms, consistent with the trends in abstention rates, the non-linear trend observed in the vote predictability of \emph{Voter Group}s is only consistent with the predictions of the reserve S-shaped functional form. 

\begin{figure}[h!]
\begin{center}
 \includegraphics[scale=0.9]{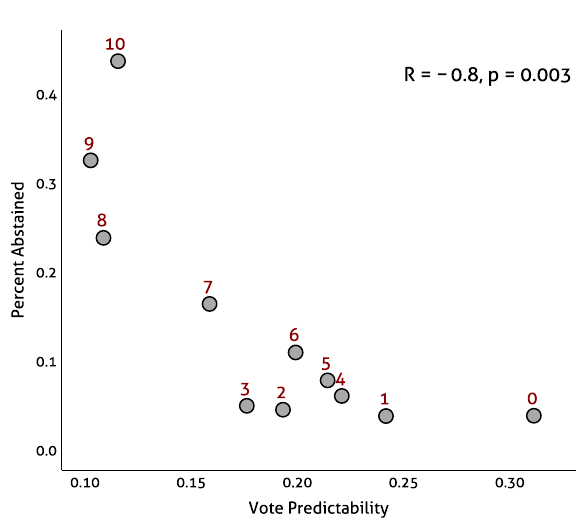}
\end{center}
 \vspace{-0.45cm}
 \caption{Correlation Between Two Measures of Voter Indifference: Abstention and Vote Unpredictability in Democrat vs. Democrat Races}
\label{fig_indiff_corr_ideoAvg}
\end{figure}

The observed correlations between the two empirical measures of indifference is further demonstrated in Figure \ref{fig_indiff_corr_ideoAvg}. The strong, negative correlation between the two at the \emph{Voter Group} level indicates that just as the rate of abstention increases, vote predictability decreases within each \emph{Voter Group}. This serves the dual purpose of providing further evidence that voter indifference is captured by both our empirical measures: vote predictability and abstention.\footnote{In the Appendix, we demonstrate that these variables are also correlated within the Democrat vs. Democrat races, as well as offices, though there are exceptions. The US House, in particular, contains outliers - Figure \ref{fig_indiff_corr_rep} of the Appendix shows this with a set of two way scatter plots of Abstention and vote predictability for each US House District. The three districts with stark anomalies are anomalous in the same way: they each have extremely high unpredictability amongst \emph{Voter Group} $0$. We see this at an extreme in two races in Los Angeles county, for which we have considerable missingness due to ballot fragmentation: District's 29 and 34.}

\begin{remark}
Voter's indifference to the candidates is observed to increase with increasing spatial distances. This observation is supported by the predictions of convex and reverse S-shaped functional forms independent of the assumptions over probabilistic voting function. It is not possible to convincingly distinguish the predictions of convex and reverse S-shaped loss functions in this setting due to the lack of information about the exact positions taken by the candidates.
\end{remark}

Though we find patterns with both measures of indifference that are consistent the predictions of the convex and the reverse S-shaped functional forms, we have yet to provide evidence that would allow us to distinguish between convex and reverse S-shaped voter loss. While the non-monotone predictability trends are only supported by the reverse S-shaped functional form, this is far from convincing evidence to distinguish it from the predictions of the convex functional form. It is possible that this is simply an empirical limitation as we do not perfectly observe the exact positions of the candidates in each race. Yet, we do observe non-monotone trending abstention rates in some races. 

For instance, Figure A.2 of the Appendix, which displays abstention by \emph{Voter Group} separately for each individual Democrat vs. Democrat race, shows a U-shaped trend for State House districts 46, 53, and 63; as well as State Senate district 33, and US House districts 34 and 44.  In \emph{Case 2}, we find corroborating evidence for these trends. 

\subsection*{Case 2: Voter Behavior in Races with Opposing Party Candidates and Varying Polarization}

We now turn to races between opposing party candidates, $c_D$ and $c_R$, where we can observe voter distance to both candidates increase simultaneously. Unlike with the same party races, our goal here is to analyze the behavior of one specific \emph{Voter Group}, who we define as `moderates', while both candidates grow more distant from them but in opposite directions. While the results are consistent with our analysis of same party races, the focus on moderate voters here enables us to further distinguish the predictions of the convex and reverse S-shaped functional forms.

We define moderates as the \emph{Voter Group} that exhibits the greatest indifference between Democratic Republican candidates using a measure of how predictable their votes are as a group. We operationalize predictability by identifying the group closest to a 50-50 split between the Democratic and Republican candidates on average across all of the federal and state level races with candidates from both parties present.  Figure A.13 of the Appendix displays the average proportion of support for the two major parties across all national and state level races by \emph{Voter Group} for both states, and indicates very clearly that in Colorado, this group is \emph{Voter Group} $5$, while in California, there is no single group that is decidedly indifferent as in Colorado. Hence, we define the moderate voters to be those in \emph{Voter Group}s $6$ \emph{and} $7$. 

Measuring candidate positions presents an empirical challenge as we do not observe variation in support for the state ballot propositions among candidates from the same party. Here, we take a more data driven approach. We base our measure of polarization on the behavior of the \emph{Voter Group}s who, by definition, most align with the two parties' positions. That is the percent support for the Democratic Party candidates among \emph{Voter Group} $0$ and that for the Republican Party candidates among the \emph{Voter Group} $n$ at the race level. We can expect these two voter groups' support for the opposing party candidates to be larger as the candidates get more polarized, as the spatial models of voting predict independent of the assumed functional form.

More formally, the polarization measure for each race $j$ is the product of the proportion of voters in \emph{Voter Group} $0$ that voted for the Democratic Party candidate, $Pr_{k=0}^j(D)=\frac{N_{k=0}^{j}(D)}{N_{k=0}^j}$, and proportion of voters in \emph{Voter Group} $n$ that voted for the Republican Party candidate. $Pr_{k=n}^j(R)=\frac{N_{k=n}^{j}(R)}{N_{k=n}^j}$. Then, the polarization level in race $j$ is given by $pol_{j}  = Pr_{k=0}^j(D) . Pr_{k=n}^j(R)$. The distribution of our candidate polarization measure is shown in the Appendix. Importantly, this measure is independent of the voting behavior of the moderate voters allowing us to examine the moderate voters' voting behavior as the opposing party candidates both move to opposite ends of the spatial spectrum, thereby increasing their spatial distances to the moderate voters. 

With these measures fully defined, we now assess effect of candidate polarization on the abstention behavior and vote predictability of the moderate voters through a series of regressions presented in Table \ref{table_polar_comb} for California and Table \ref{table_polar_comb_co} for Colorado. Here, our aim is to see whether increasing candidate polarization has a monotone or non-monotone effect on moderate voters' indifference to the opposing candidates as well as observing the direction of these effects. As discussed in the formal predictions sections, a null effect is supported by the linear loss function, a monotone decreasing effect is supported by the strict concave loss functions, a monotone increasing effect is supported by the convex loss functions and a non-monotone effect that first increases and then decreases is supported by the reverse S-shaped loss functions. 

First, we examine this relationship with a set of linear piece wise regressions, shown in the first four columns of each table. In these models, we test if there exists a threshold polarization level such that among races where polarization level below this threshold, the effect of increasing candidate polarization is not on the same direction as the effect of increasing candidate polarization among races where polarization level above this threshold. In both states and for both measures, this threshold exists. We define this threshold as the average candidate polarization within each state. The first and third columns show the effect of polarization on our two measures of indifference among moderate voters in relatively low polarization races, while columns two and four provide the same effect but in races with relatively higher level of polarization. Then, in columns five and six, we test for the same pattern but without directly imposing the threshold but instead using a quadratic model where we also include a squared term on candidate polarization, $pol_j^2$. 

The piecewise linear and quadratic regressions together allow us to gather more robust and intricate support for the U-shaped relationship between voter indifference and increasing spatial distances to both candidates. As candidate polarization increases, the moderate \emph{Voter Group}s first exhibit less indifference but with as candidate polarization continues to grow, they begin to grow increasingly indifferent to both.  Further, these patterns are consistent for both measures of indifference in both states. The average abstention rate among this group first decreases and then increases with candidate polarization, while their vote predictability first increases and then decreases. In our individual level regression analyses, presented in Tables \ref{table_polar_VG} and \ref{table_polar_co_VG}, we show that these \emph{Voter Group} level patterns also hold at the individual voter level for abstention. Though our vote predictability measure cannot be taken at the individual voter level,  abstention is by definition an individual level phenomenon. In this analysis, we use voter level fixed effects and apply the same pairwise linear and quadratic regressions over moderate voters' abstention behavior across races with varying levels of candidate polarization. 

The non-monotone U-shaped relationship between voter indifference and increasing spatial distance to both candidates that we observe in our regression analyses at both the \emph{Voter Group} and individual level is fully, and only, consistent with the predictions of the reverse S-shaped functional form. 

\begin{remark}
The observed relationship between the voter indifference and increasing spatial distances to both candidates follows a U-shape that is only supported by the reverse S-shaped loss functions. Other loss functions, convex, that can support the general trends in \emph{Case 1}  fail to match the more nuanced voting behavior captured in \emph{Case 2}. 
\end{remark}

The non-monotone trends in abstention rates as a function of candidates' spatial distances to the voter also suggests that it is the difference in utility between the two candidates, rather than the utility from the voter's most preferred candidate as in the alienation approach to abstention, that best explains observed voter behavior. While it is not possible to distinguish the predictions of the convex loss function under a framework where abstention is a function of the voter's stake in the election, and the alienation framework where abstention is a function of voter's utility from their preferred candidate, the same is not true for the alienation framework and the predictions of reverse S-shaped functional form when abstention is a function of the stakes in the election. The non-monotone trends observed in this case with regard to voter abstention rates not only corroborates the reverse S-shaped function under our approach to abstention but also provides evidence against the alienation approach to modeling abstention.

\begin{center}
\begin{table}[!h]
\caption{\normalsize Race Polarization and Indifference in California 2020}
\label{table_polar_comb}
\fontsize{10}{12}\selectfont
\begin{tabular}[t]{lccllllclll}
\toprule
 & \multicolumn{5}{c}{Linear Piecewise}&& \multicolumn{4}{c}{Quadratic}\\
 & &  && & & &  && &\\
 \cmidrule(lr){2-6} \cmidrule(lr){8-11} 
 & \multicolumn{2}{c}{Roll-off} &&    \multicolumn{2}{c}{Predictability}&&Roll-off &&\multicolumn{2}{c}{Predictability}\\
\midrule
Polarization & \num{-0.826}* & \num{0.275}* &&    \num{2.101}* &\num{-4.493}* &&\num{-8.228}*&&\multicolumn{2}{c}{\num{48.389}*}\\
 & (\num{0.094}) & (\num{0.007}) &&    (\num{0.011}) &(\num{0.004}) &&(\num{0.148})&&\multicolumn{2}{c}{(\num{0.101})}\\
 & &  &&   &&&  && &\\
Polarization$^{2}$ &  &   &&    &&&\num{4.474}*&&\multicolumn{2}{c}{\num{-27.994}*}\\
 &  &   &&    &&&(\num{0.081})&&\multicolumn{2}{c}{(\num{0.054})}\\
\midrule
No. of Observations & \num{1105693} & \num{782170}  &&    \num{267721} &\num{1620850}  &&\num{1887863}  &&\multicolumn{2}{c}{\num{1888571}}\\
 Voter Fixed Effects& Yes& Yes& & No& No& & Yes& & \multicolumn{2}{c}{No}\\
\bottomrule
 \multicolumn{3}{l}{\rule{0pt}{1em}Standard errors in parentheses.} &&\multicolumn{4}{l}{} & & &\\
 \multicolumn{3}{l}{\rule{0pt}{1em}* p $<$ 0.01} &&\multicolumn{4}{l}{} & & &\\
\end{tabular}
\end{table}

\end{center}

\begin{center}
\begin{table}[!h]
\caption{\normalsize Race Polarization and Indifference in Colorado 2020}
\fontsize{10}{12}\selectfont
\begin{tabular}[t]{lccllllclll}
\toprule
 & \multicolumn{5}{c}{Linear Piecewise}&& \multicolumn{4}{c}{Quadratic}\\
 & &  && & & &  && &\\
 \cmidrule(lr){2-6} \cmidrule(lr){8-11} 
 & \multicolumn{2}{c}{Roll-off} &&    \multicolumn{2}{c}{Predictability}&&Roll-off &&\multicolumn{2}{c}{Predictability}\\
\midrule
Polarization& \num{-0.308}* & \num{0.234}*  &&    \num{15.772}* &\num{-0.842}*  &&\num{-2.189}*&&\multicolumn{2}{c}{\num{57.710}*}\\
 & (\num{0.046}) & (\num{0.012})  &&    (\num{0.000}) &(\num{0.010})  &&(\num{0.286})&&\multicolumn{2}{c}{(\num{0.243})}\\
 & &  &&   &&&  && &\\
Polarization$^{2}$ &  &   &&    &&&\num{1.290}*&&\multicolumn{2}{c}{\num{-30.493}*}\\
 &  &   &&    &&&(\num{0.155})&&\multicolumn{2}{c}{(\num{0.131})}\\
\midrule
No. of Observations & \num{610172} & \num{716379}  &&    \num{169053} &\num{1719518} &&\num{1326551}&&\multicolumn{2}{c}{\num{1326551}}\\
 Voter Fixed Effects& Yes& Yes& & No& No& & Yes& & \multicolumn{2}{c}{No}\\
\bottomrule
 \multicolumn{3}{l}{\rule{0pt}{1em}Standard errors in parentheses.} &&\multicolumn{4}{l}{} & & &\\
 \multicolumn{3}{l}{\rule{0pt}{1em}* p $<$ 0.01} &&\multicolumn{4}{l}{} & & &\\
\end{tabular}
\label{table_polar_comb_co}
\end{table}

\end{center}

\begin{center}
\begin{table}[!h]
\caption{\normalsize Race Polarization and Roll-off Within Voter Group in California 2020}
\fontsize{10}{12}\selectfont
\centering
\begin{tabular}[t]{lccc}
\toprule
\multicolumn{1}{c}{ } & \multicolumn{2}{c}{Linear Piecewise} & \multicolumn{1}{c}{Quadratic} \\
\cmidrule(l{3pt}r{3pt}){2-3} \cmidrule(l{3pt}r{3pt}){4-4}
  & (1) & (2) & (3)\\
\midrule
Race Polarization & \num{-0.239}* & \num{0.173}* & \num{-6.118}*\\
 & (\num{0.003}) & (\num{0.001}) & (\num{0.033})\\
Race Polarization$^{2}$ &  &  & \num{3.309}*\\
 &  &  & (\num{0.018})\\
\midrule
No. of Observations & \num{169053} & \num{1719518} & \num{1888571}\\
Voter Fixed Effects & No & No & No\\
\bottomrule
\multicolumn{4}{l}{\rule{0pt}{1em}\textit{Note: } Standard errors in parentheses.}\\
\multicolumn{4}{l}{\rule{0pt}{1em}* p $<$ 0.01}\\
\end{tabular}
\label{table_polar_VG}
\end{table}

\end{center}

\begin{center}
\begin{table}[!h]
\caption{\normalsize US House Race Polarization and Roll-off Within Voter Group in Colorado 2020}
\fontsize{10}{12}\selectfont
\centering
\begin{tabular}[t]{lccc}
\toprule
\multicolumn{1}{c}{ } & \multicolumn{2}{c}{Linear Piecewise} & \multicolumn{1}{c}{Quadratic} \\
\cmidrule(l{3pt}r{3pt}){2-3} \cmidrule(l{3pt}r{3pt}){4-4}
  & (1) & (2) & (3)\\
\midrule
Race Polarization & \num{-0.308}* & \num{0.284}* & \num{-2.977}*\\
 & (\num{0.000}) & (\num{0.001}) & (\num{0.020})\\
Race Polarization$^{2}$ &  &  & \num{1.715}*\\
 &  &  & (\num{0.011})\\
\midrule
No. of Observations & \num{610172} & \num{716379} & \num{1326551}\\
Voter Fixed Effects & No & No & No\\
\bottomrule
\multicolumn{4}{l}{\rule{0pt}{1em}\textit{Note: } Standard errors in parentheses.}\\
\multicolumn{4}{l}{\rule{0pt}{1em}* p $<$ 0.01}\\
\end{tabular}
\label{table_polar_co_VG}
\end{table}

\end{center}

\section*{Discussion}

Our analysis reveals several clear patterns. First, voters appear to become more indifferent as their ideal position diverges from those of the candidates choices. Second, voter indifference is not monotonic in spatial distance, but rather increases as the position of both candidates grows sufficiently close and far from the voter. Though this may sound intuitive, it is not the pattern predicted by the most commonly used concave functional form assumptions in the formal and empirical literature. Instead, our results show that the functional form that best explains how voters assign utility to alternative candidate choices is the reserve S-shaped functional form. This also happens to be the function whose predictions are least sensitive to specific assumptions over the dimensionality of policy space and the probabilistic voting functions, which can in certain settings mask the shortcomings of the assumptions over the functional form of voter utility. 

Our findings have important implications for almost all aspects of electoral competition, including, but not limited to, voters' ballot choices, turnout decisions, the strategic platform choices by the parties and their candidates, vote buying and allocation of campaign spending. In this section, we discuss our findings' robustness to alternative explanations and the implications of our findings on the formal and empirical models of electoral politics.  

\textbf{Alternative explanations} 
One may contend that our results arise from phenomena other than voter preferences such as asymmetric information across \emph{Voter Groups}. Indeed, it may be the case that voters do not get more indifferent to the candidates as candidate positions diverge from that of their own but, instead, the voter cannot distinguish these candidates' positions due to a lack of information. This argument applies especially to \emph{Case 1} of our analysis when candidates from the same party compete. In this setting, the story behind this argument would be as follows: Voters belonging to more conservative \emph{Voter Group}s attain their information from news sources that provide less coverage on contests that do not feature a Republican candidate, and are thereby less informed about the two Democratic Party candidates' than the voters in more liberal \emph{Voter Group}s. 

This story hardly applies to Case $2$ which features races between opposing party candidates as we have no reason to believe that races between more polarized candidates receive less media coverage, yet, the results of Case $2$ are consistent with those in Case $1$. Though, let us consider the alternative explanation to Case $1$ more carefully. Attaining information on candidates is costly to voters, mostly through its opportunity cost. Then, the voters who we should expect to pay this information cost should be those for whom expected returns from information on the specific races is higher. With this rationale, for the alternative information information story to hold in races between two Democratic candidates, the voters whose utility is expected to be least affected by the potential differences in the two Democratic candidate platforms should be the more conservative voters. Hence, even if it were the case that the conservative voters are those with least information about the candidates, this is the result of their indifference which can be supported by the convex and reverse-sigmoid loss functions.

\textbf{Implications for Formal Models of Electoral Politics}

Formal models of electoral competition primarily focus on strategic platform choices by competing political parties and their candidates. Regarding voter utility, the most commonly adapted assumption in this literature is that voter utility follows concave functional form, in the form of quadratic and linear loss functions, given the mathematical convenience they offers. It is generally the hope and the assertion in these studies that the assumption over concave loss functions are without loss of generality. Yet, it has been shown that even in the most rudimentary models, when abstention is accounted for or a probabilistic voting approach is taken, the concave assumption is in fact not without a loss of generality \citep{valasek2012turnout, callander2007turnout, zakharov2008}.

To see this, consider a basic electoral competition model with two candidates, $c_1$ and $c_2$, and a continuum of voters distributed across a uni-dimensional policy space. In the first stage, candidates simultaneously choose policy platforms and in the second stage, voters cast their votes for a candidate or abstain. As there will be multiple, and infinitely many with continuum of voters, equilibria in the voting stage independent of the candidates' platform choices. With further assumptions, each voter's decision rule is ensured to be voting for the candidate providing the voter sufficiently large utility that is $c_1$ if $|u_i(c_1)-u(c_2)|>c$. 

If the voter's preferences are best captured by a concave loss function, in a deterministic setting, both candidates are incentivized to converge to the median voter's position, independent of the distribution of voters along the policy space, policy and office motivations of the candidates and possibility of voter abstention. Under the probabilistic voting framework, incentives for divergence from the median voter's position towards the candidate's preferred position are created only if the candidate is sufficiently policy motivated such that the gains from the policy outweighs the diminishing electoral prospect.\footnote{In citizen candidate models when it is assumed that electoral platforms only serve as cheap talk, voter have perfect information of candidate's preferred policies and know that the policy to be implemented will be the candidate's preferred policy instead of the electoral platform, both convergence and limited divergence results can be supported.} These results emerge under concave loss assumptions since convergence to the median voter's ideal position does not cost the candidates votes from voters whose ideal positions are sufficiently distant to both candidates' positions. In a uni-dimensional policy space, these are the voters with extreme positions. Hence, the competition between the parties and their candidates is over the moderate voter or voters who are positioned around the median. 

If, as our results suggest, the reverse S-shaped loss function is adapted in this body of work, converging towards the median voter and competition over these voters will cost the candidates the support of the extreme voters. Then, in models of probabilistic voting and deterministic voting models accounting for voter abstention such that $c>0$, a candidate may face incentives to diverge from the median voter's preferred position purely to increase their electoral prospects. These incentives becomes apparent as the electorate become more polarized which can be visualized with a bi-modal distribution of voters' ideal policy positions. In this case, even if the candidates are purely office motivated, their platforms will respond to voter polarization if voter utility is given a reverse S-shaped functional form. With regards to the candidates' platform choices in equilibrium, we can observe pure strategy equilibrate where candidate platforms converge but not necessarily to the median voter's position, where candidate platforms diverge as well as cases where there exists no pure strategy equilibrium depending on the distribution of voters' ideal policy positions. We cover these cases in the Appendix and an analysis of the simple electoral competition game under different functional form assumptions is also well covered in \citet{valasek2012turnout}.

Even in a primary contest, strategic voting incentives among extreme voters leads to convergence if we assume voter utility follows a concave function. Adopting the reverse S-shaped loss function also has implications for voters' strategic voting behavior and consequently, the types of candidates who are expected to run in the primary election. Under this assumption, there will be large regions of the policy space over which voters are risk-seeking rather than risk-averse. For instance, faced with a choice between a moderate and a more extreme candidate in the primary election, voters with more extreme policy preferences would not necessarily have the incentive to strategically vote for the moderate candidate to increase their party's chances in the general elections. That is, these voters may prefer their party put forward a more extreme candidate even though it could decrease the party's overall electoral chances. This can in turn dis-incentivize moderate candidates and incentivize more extreme candidates to run, especially when the electorate is more polarized. These implications can be extended to electoral settings with multiple candidates and parties, as well as to explain the allure of single-issue parties for some voters, especially those whose preferences are not represented by the mainstream parties without relying on assumptions over sincere voting.

Formal models in the literature on vote buying in legislatures and campaign spending concern the allocation problem by the parties and interest groups mainly over which groups of legislators and voters to target. The legislators in legislative vote buying and the voters in targeted campaign spending are those whose votes require the least spending for an equal sized shift. In these models, when voter or legislator utility is assumed to be concave, the target voters are the more moderate groups, as they are the only indifferent voter block. If, on the other hand, the assumptions over legislator and voter preferences are taken to be best represented by the reverse S-shaped loss functions, as our findings suggest, the cheapest to target voter block will include more than one group. In a uni-dimensional policy space, it will include extreme voter groups, and in a multi-dimensional policy space, those whose ideal points are further from the choices on the ballot. Whether these other voter groups are indeed more likely to be targeted by special interest campaign spending and parties legislative vote buying than moderate voters is itself an empirical question. Nevertheless, the predictions from these models can undoubtedly be enhanced by representing voter utility with the functional form that is empirically validated -  the reverse S-shaped function.

\textbf{Implications for Empirical Models of Electoral Politics}
Our findings are most consequential for two main branches of the empirical literature: (1) research applying or generating ideal point estimates from observed voting behavior, and (2) studies focused on inferring voter behavior from their stated or revealed preferences on issues and candidate attributes such as race and gender.

The large literature on ideal point estimators focuses primarily on legislatures, with the aim of identifying legislators' preferred policy positions given their roll-call records or campaign contributions. These ideal point estimators assume a policy space composed of dimensions representing each legislation. Based on the utility maximizing legislators' roll-call votes, a cutoff point is assigned to each legislation and an ideal policy position is assigned to each legislator. A straightforward difference between assuming a concave and reverse S-shaped loss function appears when we consider a legislator exhibiting indifferent voting behavior; under a concave loss function, this legislator would be positioned as a moderate but assuming a reverse S-shaped loss function distinguishes the moderate from the extreme positions. This in turn can also help us better understand how polarized the legislature is, and the extent to which factors such as party discipline and bargaining processes are at play in legislative voting behavior.

Further, these studies \emph{infer} the number of policy dimensions required to explain voting behavior. That is, they infer the minimum number of dimensions required to predict the voting behavior of both the legislators and voters. Since the actual positioning of specific legislators and voters will depend on the assumed functional form of their utility, it is also possible that the dimensions required to explain voting behavior will change \emph{with} the functional form. While the actual ideal point estimates will depend on the real distribution of legislators and voters, it is important to understand the circumstances under which we can expect these estimators to generate divergent predictions.

The most widely adopted ideal point estimators include NOMINATE, DIME, IDEAL, and PCA, which each adopt different functional form assumptions over legislator utility. This is despite that existing research has established robust empirical evidence for the utility function that best describes legislator behavior \citep{carroll2013structure}. It is clear that these findings have simply not been adopted by this literature. The results we present in this article for the shape of voter utility are in full coherence with evidence presented in \citet{carroll2013structure} for the reverse S-shaped loss functions. 

It is also important to consider the empirical research that seeks to infer voter policy preferences from behavior and vice versa. A question this literature aims to answer is the extent to which specific issue domains, such as economic or social, affect voter behavior. These studies measure voter preferences either through survey data or with a causal inferential approach that relies on shocks believed to increase the salience of certain issues, such as natural disasters. Conceptually, these studies adopt an issue based approach complementing a spatial component to measure voter utility. That is, a voter's utility from issue $n$ is evaluated separately from other dimensions which are generally controlled for. If voter $i$'s utility from issue $n$ is $u({x_{i_n}}, x_{c_n})$, where the voter's preferred position on the issue is $x_{i_n}$ and the position candidate $c$ takes or is attributed to $x_{c_n}$, voter $i$'s utility from candidate $c$ will correspond to $u(x_{i_{n}},x_{c_{n}})+u(x_{i_{n'}},x_{c_{n'}})$. In this formulation, the issue of interest, $n$, is treated in the issue based approach and the positions taken on other issues, represented by $n'$, are treated as a separate component. 

As a voter's choice on the ballot will be a function of the difference in utility between the specific choices, a linear regression model aiming to capture the effect of issue $n$ the voter's candidate choices will embed the assumption of a linear or a quadratic loss function over their utility, as these functional form assumptions predict a linear relationship between the voter's position and their net utility from the candidates.\footnote{While this holds only in a uni-dimensional setting under the linear functional form assumptions, it holds in both uni-and multi- dimensional settings under the quadratic loss function.} If we assume a reverse S-shaped form loss function, we of  course require a different model to assess the true impact of preferences on choices. This issue is not limited to linear regression - it includes all models that assume a monotone relationship between voter preferences and their choices. For example, consider a study that aims to assess the effect of voters' economic policy preferences on their decision to vote for the Democratic or Republican candidate. The conventional approach would be to regress the Democratic candidate vote on a measure of how liberal the voter's economic policy preferences are. An analysis assuming a monotone relationship between these two variables will overestimate the effect for some voter groups and underestimate others. That is, a reverse S-shaped loss function would suggest that the more liberal their economic policy preferences, the more likely it is that a voter chooses the Democratic party. But only up to a point. This relationship will dissipate for voter groups whose preferences are more liberal than the Democratic candidates. So a monotone regression model will overestimate the effect of policy preference for the latter group of voters and underestimate it for the former group.

Taken together, establishing the functional form that best captures voter preference relations has important implications across a range of approaches to the study of voter behavior and electoral politics, and independent of the specific methods researchers employ. Our findings using individual level voter data corroborate those found in existing research on legislatures \citep{carroll2013structure}, together suggesting that the reverse S-shaped loss functions best capture the behavior of both legislators and voters. As discussed, for both the empirical and formal theoretical bodies of research, adopting the assumption that is empirically validated will enable researchers to generate more nuanced and coherent predictions, as well as interpretations of their results. One can also hope that such a unifying baseline assumption may increase collaboration between these two groups of researchers, as should be in the science of politics.

\bigskip
\newcommand\EatDot[1]{}

\bibliographystyle{apsr}
\bibliography{utility_bib}

\newpage

\renewcommand{\thetable}{A.\arabic{table}}
\renewcommand{\thefigure}{A.\arabic{figure}}

\appendix
\clearpage

\FloatBarrier

\setcounter{table}{0}
\setcounter{figure}{0}
\setcounter{page}{1}

\thispagestyle{empty}

\begin{center}
\textbf{\LARGE Appendix}
\end{center}

\tableofcontents
\newpage

\label{sec:append}

\section{Formulation of Voter Payoffs}

In the context of voting in a race between the two candidates, $c_1$ and $c_2$, the voter $i$'s choice over abstaining or voting and if so which candidate to vote for can be rationalized through a pivotal voting or expressive voting frameworks. Formulating voter payoffs from voting for each candidate and abstaining through these approaches differ mainly through the voter's payoff from abstaining whether is it a fixed payoff independent of the candidate options or a function of the candidate positions and the voter's payoff from these candidates. 

If we take the pivotal voting framework, voter $i$'s payoff from voting for the candidate $c_1$, voting for the candidate $c_2$, and abstaining, denoted by $U_i(.)$, can be expressed as:

$U_i(c_1)=u(\delta_{i,c_1})-c$
$U_i(c_2)=u(\delta_{i,c_2})-c$
$U_i(A)=\frac{u(\delta_{i,c_1})+u(\delta_{i,c_2})}{2}$

In this approach, the voter undertakes the cost of voting if she casts a vote for either candidate, $c$, and receives payoff of $u(\delta_{i,c_1})$ in the event that candidate $c_1$ wins the election. As the voter is considered as the pivotal voter, at least regarding her vote choice, in case of abstention, the voter does not undertake the cost from voting and in exceptions receives a payoff of a lottery where both candidates have equal chance of getting elected. From a behavioral approach, it is also possible think of voter's expectation from abstaining as the worse of the two candidates getting elected such that $U_i(A)=\min\{u(\delta_{i,c_1}),u(\delta_{i,c_2})\}$. Though, sticking to the pivotal voting approach, the voter $i$ will vote for candidate $c_1$ if  $u(\delta_{i,c_1})-u(\delta_{i,c_2})>2c$, vote for candidate $c_2$ if $u(\delta_{i,c_2})-u(\delta_{i,c_1})>2c$, and abstain from the race if $|u(\delta_{i,c_1})-u(\delta_{i,c_2})|>2c$. Given the this decision rule, voter $i$'s indifference to the candidates can be captured by the term $-|u(\delta_{i,c_1})-u(\delta_{i,c_2})|$ higher values of which correspond to growing voter indifference between the candidate choices. 

In models of probabilistic voting, our representative voter $i$'s payoff from voting for candidate $c_1$, voting for candidate $c_2$ and abstaining will be:

$U_i(c_1)=u(\delta_{i,c_1})-c+\epsilon_1$
$U_i(c_2)=u(\delta_{i,c_2})-c+\epsilon_2$
$U_i(A)=\frac{u(\delta_{i,c_1})+u(\delta_{i,c_2})}{2}$

In this payoff formulation $\epsilon_1$ and $\epsilon_2$ are stochastic terms with mean zero distribution. Then, the probability that voter $i$ votes for candidate $c_1$ is given by the probabilistic voting function:
\begin{align*}
Pr_i(c_1)&=Pr\left(U_i(c_1)>U_i(c_2) \land U_i(c_1)>U_i(A)\right)\\
&=Pr\left(U_i(c_1)>U_i(c_2) | U_i(c_1)>U_i(A)\right).Pr(U_i(c_1)>U_i(A))\\
&=Pr(U_i(c_1)>U_i(A))\\
&=1-\Phi_{\epsilon_1-\epsilon_2}(u(\delta_{i,c_2})-u(\delta_{i,c_1})+2c)
\end{align*}
Similarly, probability that voter $i$ votes for candidate $c_2$ is given by $Pr_i(c_2)=\Phi_{\epsilon_1-\epsilon_2}(u(\delta_{i,c_2})-u(\delta_{i,c_1})-2c)$ where $\Phi_{\epsilon_1-\epsilon_2}(.)$ is the cumulative distribution function over $\epsilon_1-\epsilon_2$. Assuming $\epsilon_1$ and $\epsilon_2$ to have a uniform distribution leads to a linear probabilistic voting function, while assuming $\epsilon$ to have a normal or logistic distributed leads to a reverse S-shaped probabilistic voting function as in the logit and probit models. Then, voter $i$ is expected to abstain with probability $Pr(A)=1-Pr_i(c_1)-Pr_i(c_2)= \Phi_{\epsilon_1-\epsilon_2}(u(\delta_{i,c_2})-u(\delta_{i,c_1})+2c)-\Phi_{\epsilon_1-\epsilon_2}(u(\delta_{i,c_2})-u(\delta_{i,c_1})-2c)$ and as in the deterministic case, the probability that the voter abstains in a race between candidates $c_1$ and $c_2$ increases as the voter's indifference to the candidates, which is captured by the term $-|u(\delta_{i,c_1})-u(\delta_{i,c_2})|$, increases. Realize that in this formulation, if the voter prefers to vote for candidate $c_1$ over abstaining in the deterministic setting, she prefers abstaining over voting for candidate $c_2$. Further, in the probabilistic setting, decreasing voter's payoff from her less prefer candidate decreases not only the probability of voting for this candidate but of that of abstention. These are not the cases if the voter is assigned a constant payoff from abstaining and independent payoffs from voting for either candidate as in some expressive voting models.

Now let us consider a basic payoff structure in an expressive voting model where the voter's payoff from voting for the candidates remains as above but there is a constant payoff $a$ from abstention independent of the candidate choices. Under this payoff structure, in the deterministic setting, it is possible for a voter for prefer $c_1$ to $c_2$ and still prefer voting for candidate $c_2$ to abstaining in the race. That is, the voter is better off contributing to the victory of her worst candidate option than abstaining in the race. Further, in the probabilistic setting, decreasing the voter's payoff from their less preferred candidate not only increases the voter's probability of voting for her preferred candidate but also probability of abstention. As earlier mentioned, this formulation leads to expectations that a liberal voter is more likely to abstain in a race between Joe Biden and Donald Trump than in a race between Joe Biden and Joe Biden even though the stakes in the former election is much higher. That is, this formulation generates results consistent with the alienation approach to abstention and also generates further predictions. 

It is possible to have fixed payoff from abstention in an expressive voting setting and generate equivalent results to the pivotal voter framework. For this, voter's payoff from voting for either candidate needs to be a function of the stakes at the election. That is, payoff from voting candidate $c_1$ is not only a function of this candidate's position but also the opponent's such as $U_i(c_1)=u(\delta_{i,c_1})-u(\delta_{i,c_2})-c+\epsilon_1$. In this formulation, without making specific assumptions over voter's expectations in case of abstention, counter-intuitive implications of the expressive voting framework can be avoided. 

With regards to abstention, while it is not possible to distinguish the predictions of the convex loss function under a framework where abstention is a function of the stakes in the election and the alienation framework where abstention is a function of voter's utility from their preferred candidate, the same is not true for the alienation framework and the predictions of reverse S-shaped functional form when abstention is a function of the stakes in the election. The non-monotone trends observed in \emph{Case 2} with regards to voter abstention rates not only corroborates reverse S-shaped functional form under our approach to abstention but also provides evidence against the alienation approach to abstention.

\section {Spatial Models in Multi-Dimensional Policy Space}

If policy space is taken to be multi-dimensional, our representative voter's ideal policy position and the positions taken by the candidates will take the vector form. Then, voter $i$'s utility from candidate $c$ is expressed as $u_i(c)=u(\delta_{i,c})$ where $\delta_{i,c}=|\vec{x_i}-\vec{x_c}|$ is the spatial distance between the voter's and the candidate's position vectors, and function $u(.)$ maps the spatial distances into voter's utility from the candidate. Unlike in the uni-dimensional context, in the multi-dimensional policy space, the effect of shifts in the voter's position vector, $\Delta x_i$ as in \emph{Case 1}, or in candidates' position vectors, $\Delta pol$ as in \emph{Case 2}, on voter's spatial distance to the candidates is not linear. The representations of these shifts in multi-dimensional policy space is shown in Figure.... That is, as the voter's spatial distance to both candidates, $c_1$ and $c_2$, increases, the difference between the voter's spatial distance to the two candidates, $|\delta_{i,c_1}-\delta_{i,c_2}|$, decreases, formally, $\frac{\delta \delta_{i,c}}{\delta^2 x_i}<0$. 

As the position shifts away from both candidates do not have equisized effects on the voter's spatial distance to both candidates, their effect on voter's indifference to candidates, $|u_i(\delta_{i,c_1})-u_i(i,c_2)|$ under some functional form assumptions are also different from those in the uni-dimensional policy space. The functional form assumptions that generate different predictions over voter's  indifference to the candidates as the approach to dimensionality of policy space changes are the linear and quadratic functional forms. The linear utility functions generate predictions analogous to those of the convex functional form. That is, as the voter's spatial distance to both candidates increases, the voter monotonously gets more indifferent to the candidate choices. Whereas, the quadratic functional form predicts that the voter's indifference to the candidates remain unchanged as the spatial distances to both candidates increase in the multi-dimensional setting. 

Recalling our findings for \emph{Case 1}, the functional form assumptions, predictions of which support the results in a multi-dimensional policy setting are the linear, convex and reverse S-shaped functional forms. Regarding \emph{Case 2}, on the other hand, the non-monotone trends regarding voter indifference can only be supported by the reverse S-shaped functional form. Overall, the functional form, whose predictions are immune to the approach to dimensionality of policy space and can support the findings in our cases is the reverse S-shaped functional form.

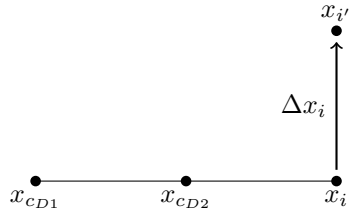
\begin{figure}[h!]
\begin{center}
\begin{tikzpicture}
    \coordinate (y) at (0,3);
    \coordinate (xCD1) at (0,0);
    \coordinate (xiP) at (4,2);
    \coordinate (xi) at (4,0);
    \coordinate (xCD2) at (2,0);
    \coordinate (x) at (4,0);
    \coordinate (x1) at (4,0.15);
    \coordinate (x2) at (4,1.85);

    \draw (xCD1) -- (x);

    \draw[->, thick] (x1) -- (x2) node[midway, left] {$\Delta{x_i}$};

    \fill (xiP) circle (2pt) node[above] {$x_{i'}$};
    \fill (xi) circle (2pt) node[below] {$x_i$};
    \fill (xCD1) circle (2pt) node[below] {$x_{c_{D1}}$};
    \fill (xCD2) circle (2pt) node[below] {$x_{c_{D2}}$};

\end{tikzpicture}




\end{center}
\caption{Illustration of \emph{Case 1} in multi-dimensional policy space where voter $i$'s position vector moves by $\Delta x_i$ to $x_{i'}$ and distances from those of both Democratic Party candidates, $c_{D1}$ and $c_{D_2}$.}
\label{appendix_tikz}
\end{figure}

\begin{figure}[h!]
\begin{center}
\begin{tikzpicture}
    \coordinate (xCD) at (0,0);
    \coordinate (xCDp) at (0,2);
    \coordinate (xi) at (2.3,0);
    \coordinate (xCR) at (4,0);
    \coordinate (xCRp) at (4,-2);
    \coordinate (x) at (4,0);
    \coordinate (x1) at (0,0.15);
    \coordinate (x2) at (0,1.85);
    \coordinate (y1) at (4,-0.15);
    \coordinate (y2) at (4,-1.85);

    \draw (xCD) -- (x);

    \draw[->, thick] (x1) -- (x2) node[midway, left] {$\Delta_{pol}$}; 
    \draw[->, thick] (y1) -- (y2) node[midway, right] {$\Delta_{pol}$};

    \fill (xCDp) circle (2pt) node[above] {$x_{c_{D'}}$};
    \fill (xCR) circle (2pt) node[above] {$x_{c_R}$};
    \fill (xCD) circle (2pt) node[below] {$x_{c_D}$};
    \fill (xi) circle (2pt) node[below] {$x_i$};
    \fill (xCRp) circle (2pt) node[below] {$x_{c_{R'}}$};
\end{tikzpicture}

    

\end{center}
\caption{Illustration of \emph{Case 2} in multi-dimensional policy space where positions taken by the opposing party candidates, $c_{D}$ and $c_{R}$, get more polarized, moving candidates to position vectors $x_{c_{D'}}$ and $x_{c_{R'}}$ and away from the moderate voter.}
\label{appendix_tikz2}
\end{figure}
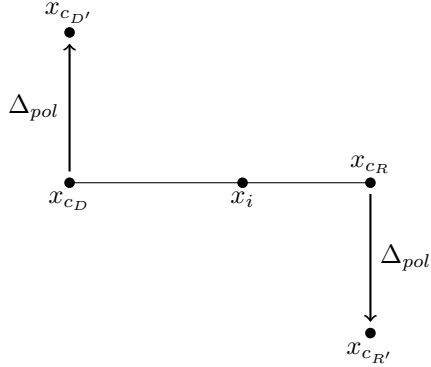

\section{Model of Electoral Competition}

Let us consider a basic model of elections where there are two office motivated candidates $c_1$ and $c_2$ and a continuum of voters whose ideal policy positions are distributed on a single policy dimension. The game unfolds in two stages, in stage 1, the candidates choose a policy platform simultaneously and in state 2, election occurs and the voters choose whether to vote for candidate 1, vote for candidate 2 or abstain. As candidates are office seekers, it can be assumed that if elected they receive a position office rent $R$, and if not, they receive a payoff $0$. If a candidate gets elected, they implement the policy platform they chose in the first stage. Voters' utility is, as generally defined, a function of their ideal policy and the platform of the winning candidate. Such that, as a general decision rule in the spatial models, we assume that the voter votes for candidate $c_1$ if $u(\delta_{i,c_1})-u(\delta_{i,c_2})>c$ and for candidate $c_2$ if $u(\delta_{i,c_2})-u(\delta_{i,c_1})>c$ where $u(.)$ is the voter's political utility function with a specific functional form which we will only consider to be concave as generally adapted in the literature and reserve-sigmoid as our analysis show to best capture voter preferences. Positions $x_i$, $x_{c_1}$ and $x_{c_2}$ represent voter $i$'s ideal policy and the candidates' platforms, respectively. The cost parameter, $c$, represents the threshold such that if the voter's net utility from a candidate is above this threshold, the voter votes for this candidate. Let us go through the equilibrium analysis under the concave loss function and the reverse S-shaped loss function. For this heuristic exercise, we will be only considering symmetric distributions of voters across the policy dimension as our aim is to only show that the convergence and existence of stable equilibrium are not general results under the reverse-sigmoid loss function.

\subsection{ Concave Loss Function}

Given the voters' decision rule, candidates whose pure motivation is to win the election will always have a weakly dominating strategy to move their platforms to the Condorcet winner position. Then, the question is whether there is a Condorcet winner policy platform under concave loss function assumption. There is a Condorcet winner platform and this is the median voter's ideal policy. To see this, let us denote the median voter's ideal policy as $x_{m}$ and assume that candidate $c_1$'s platform is positioned at $x_m$. Without loss of generality, let us consider any platform candidate $c_2$ can choose such that $x_{c_2}>x_m$. Given the concave loss function assumption, the voters supporting candidate $c_1$ will be those whose ideal positions satisfy $x_i<\frac{x_{c_1}+x_{c_2}}{2}-t$ such that $u(\frac{x_{c_1}+x_{c_2}}{2}-t-x_{c_1})-u(\frac{x_{c_1}+x_{c_2}}{2}-t-x_{c_2})=c$. And, the voters supporting candidate $c_2$ will be those with ideal policy positions $x_i>\frac{x_{c_1}+x_{c_2}}{2}+t$. The threshold $t$ is a function of the cost $c$ and the concavity of the utility function $u(.)$. If the cost is $c=0$, voters to the left of the midpoint of the two candidate platforms will be voting for $c_1$ and the rest for candidate $c_2$. As the cost increases, so does the range of voters who abstain. Though, independent of the value of $c$, the voters who abstain will be those with ideal policy positions sufficiently close to both candidate platforms. Further, since the voters supporting the opposing candidates with weakest preferences will be equidistant from the midpoint of the two candidates' platforms, there is not policy position candidate $c_2$ can take and defeat or tie candidate $c_1$ when $c_1$ is positioned at the median voter's ideal policy. Then, the best candidate $c_2$ can do is also position at $x_m$ and tie. This results of convergence also holds if, instead of a deterministic voting model, a probabilistic voting model is adapted.

\subsection{Reverse S-shaped Loss Function}

Now, let us consider the same game with voter utility functions defined as reverse S-shaped. Without loss of generality, let us assume that candidate $c_1$ has a platform positioned at $x_m$. Without loss of generality, let us consider any platform candidate $c_2$ can choose such that $x_{c_2}>x_m$. Under the deterministic model, if the voters have reverse S-shaped political utility functions, the condition that needs to be satisfied for voter $i$ to vote for candidate $c_1$ is $\frac{x_{c_1}+x_{c_2}}{2}-t_2<i<\frac{x_{c_1}+x_{c_2}}{2}-t_1$ and the voter will be voting for candidate $c_2$, if $\frac{x_{c_1}+x_{c_2}}{2}+t_2<i<\frac{x_{c_1}+x_{c_2}}{2}+t_1$ where $t_2>t_1>0$. Realize that, the candidate positioned at the median voter's ideal positions, $c_1$, does not receive the support of the extreme voters necessarily under the reverse S-shaped loss function assumption. Recall that, these voters who have ideal policy positions furthest from the median voter's ideal position were the voters with strongest preference for the candidate positioned spatially closer to them under the concave loss function assumption. Further, under the reverse S-shaped loss function assumption, the median voter's ideal policy position is not necessarily the Condorcet winner. Depending on the distribution of voters' ideal policy positions, candidate $c_2$ can defeat candidate $c_1$, especially when this distribution is not uni-modal or when the uni-modal distribution is not symmetric across the policy space. 

Then, in cases the median voter's ideal policy position is not the Condorcet winner. Yet, does there exist an equilibrium in pure strategies? The answer is not necessarily. If we take a population of voters whose ideal policy positions have a bimodal distribution and a positive valued $c$, it is possible that a policy platform that is not the median voter's position and also not defeatable by any other policy position. Then, the opposing candidate can either take a position that is symmetrical across the median voter, let us call these positions $x_{R}$ and $x_{L}$, or take the exact same position. In either case, the candidates will tie in the election in equilibrium. In the former case, the two candidates appeal to the voters with opposing policy preferences, we refer to these as $R$ for right and $L$ for left leaning voters. In the latter case, the equilibrium platform choices are not divergent from each other such that the candidates compete over the same left or right leaning voters, but the equilibrium platform choices are divergent from the median voter's ideal policy positions. Callander and
Wilson (2007) eliminate this equilibrium by allowing third candidate entry and Valasek (2012) by assuming the two candidates do not take the same policy position. Though it is important to acknowledge that even if the candidate platforms converge in these equilibria, they are not at the median voter's ideal policy position.

Under certain distributions of voters' ideal policy positions and the $c$ parameter, it is also possible that there does not exist an equilibrium in pure strategies. If the distribution is such that the two opposing platforms $x_{R}$ and $x_{L}$ lead to a tie between the two candidates but there exists a platform, $x'_{L}$ that appeals to the left leaning voters that defeat platform $x_{L}$ but loses to $x_{R}$, the game does not have a Condorcet winner platform. A sample distribution under which the disequilibirum result can hold is illustrated in the figure below. If a probabilistic voting approach is taken, this result can hold even if $c=0$. Though lack of stable equilibrium can be disheartening to some, this result already exists when the policy space is considered to be multi-dimensional and the takeaway from this simple model that the candidates have incentives to respond to the distribution of voter's ideal policy positions and choose platforms different than the median voter's ideal positions still holds and is still an important finding on the path of understanding strategic interactions between the parties and their candidates.

\begin{figure}
		\centering
		    \begin{tikzpicture}
     \coordinate (x) at (0,0);
    \coordinate (x1) at (12,0);
    \coordinate (x3) at (3,6);
    \coordinate (x4) at (3,0);
    \coordinate (x5) at (9,6);
    \coordinate (x6) at (9,0);
    \coordinate (x7) at (2.6,6);
    \coordinate (x8) at (2.6,0);
    \coordinate (x10) at (6,1.5);
    \coordinate (x9) at (6,0);

    \draw (x) -- (x1);
    \draw[dashed] (x3) -- (x4);
    \draw[dashed] (x5) -- (x6);
    \draw[dotted] (x7) -- (x8);
    \draw[dotted] (x9) -- (x10);
    
    \draw  (3,6) parabola (0,5);
    \draw  (3,6) parabola (4.5,4);
    \draw  (6,1.5) parabola (4.5,4);
    \draw  (6,1.5) parabola (7.5,4);
    \draw  (9,6) parabola (7.5,4);
    \draw  (9,6) parabola (12,5);
    \filldraw[black] (3,0) circle (2pt) node[anchor=north]{$x_{L}$};
    \filldraw[black] (9,0) circle (2pt) node[anchor=north]{$x_{R}$};
    \filldraw[grey] (2.6,0) circle (2pt) node[anchor=north]{$x'_{L}$};
    \filldraw[grey] (6,0) circle (2pt) node[anchor=north]{$x_m$};
    
\end{tikzpicture}
		\caption{\small Here illustrated a bimodal distribution of voters' ideal policies across a uni-dimensional policy line. If voters' utility over policy platforms is represented through concave loss functions, the policy platform at the median voter's ideal policy $m$ defeats alternative platforms ${x'_{L}}$, $x_{L}$ and $x_{R}$. If voter's utility is represented as reverse S-shaped loss function, there exists values of $c$ such that all three alternative platforms defeat a policy platform positioned at $x_m$. Further, it is possible that policy platforms $x_{L}$ and $x_{R}$ tie, policy platform $x'_{L}$ defeat $x_{L}$ and policy platform $x_{R}$ defeat $x'_{L}$. In this case, there is no platform that is the Condorcet winner. Should a probabilistic voting approach taken, this result can hold without the assumption over a positive valued $c$.  }    
		\label{fig:flowchartwear}
	\end{figure}
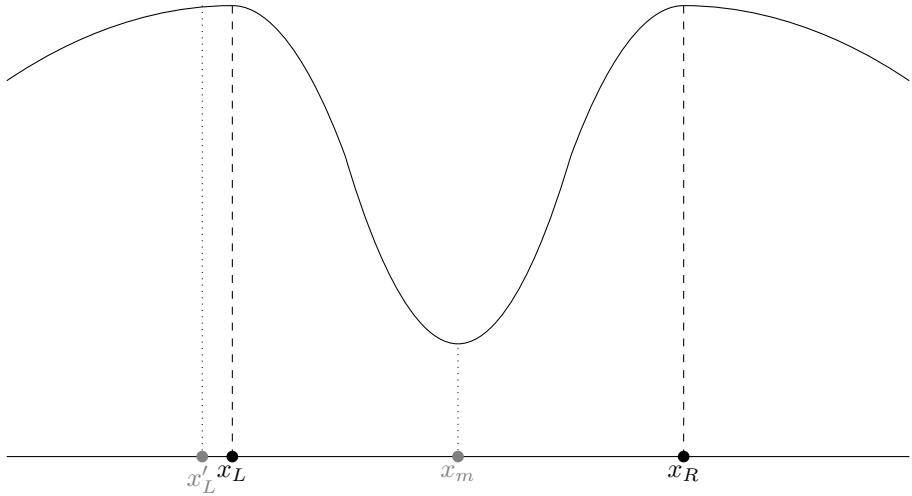

\section{Implications for Other Formulations} 

We now address the implications of our findings for other formulations of voters' political utility functions. As outlined in the \hyperref[sec:form_pred]{Formal Predictions Section}, the two primary alternatives to the spatial approach are the issue-based and directional models. We will consider whether our empirical findings can be supported by these approaches to formulating voter utilities and if so, under which functional form assumption their predictions better match the observed voter behavior. We argue that our indifference results from \emph{Case 1} and \emph{Case 2} are supported by the issue based models if the reverse S-shaped functional form is assumed and the directional voting models fall short in supporting the observations in \emph{Case 2}.  

Under the issue based approach, utility from each policy issue contributes separately and additively to the voter's utility from a candidate. In order to replicate our analysis under the spatial model using multiple \emph{Voter Groups} for singular issues, we would need a discrete measure of voter positions on these issues. The binary structure of our data on voter preferences for each policy dimension prevents us from achieving this. Though we can derive the functional form assumption that generates predictions under an issue based model that best match those of the reverse S-shaped loss functions under the spatial model. 

The spatial and issue-based approaches are analogous on a single policy dimension but this is not necessarily the case for other functional form assumptions if the policy space is modeled in multiple dimensions. To see this distinction, consider a policy space consisting of two continuous policy dimensions, $x_A=[0,1]$ and $x_B=[0,1]$, a voter having a position vector $\vec{x}_i=({x}_{i_A},{x}_{i_B})$, a candidate with platform $\vec{x}_c=(x_{c_A},x_{c_B})$. While under the spatial framework, the voter's utility from candidate $c$ is a function of their position vectors' distance, $u(|\vec{x}_i-\vec{x}_c|)$, with the issue based approach, the voter's utility from this candidate is a function of voter's utility from the candidate's position from each dimension, $u(|\vec{x}_{i_1}-\vec{x}_{c_1}|)+u(|\vec{x}_{i_2}-\vec{x}_{c_2}|)$. To uncover the functional form assumptions' predictions over voter indifference between candidates, let us assume, without loss of generality, that our representative voter $i$ has a position vector $\vec{x_i}=(1,1)$. Indifference curves for this voter under each functional form assumption and the two approaches to formulating voter utilities, spatial and issue-based, are shown in Table \ref{fig_indiff_curves}.

Under the spatial framework, this voter's indifference curve will be convex for all functional form assumptions, referred to as convex preferences. The only difference between the predictions under each functional form is the level of separation between the indifference curves. A larger separation is indicative of higher indifference, such that the voter is indifferent between the candidates who are positioned in the region between indifference curves. Separation between indifference curves will shrink at lower levels of utility under a concave loss function, remain unaffected across different utility levels under a linear loss function, grow at lower levels of utility under a convex loss function and, under reverse S-shaped loss functions, level of separation will be first shrinking and then growing as the voter's utility levels decrease. 

If voter utility is formulated under the issue based approach, voter $i$'s indifference curves will remain convex across the entire policy space if and only if the voter's utility is concave. If voter utility is shaped linearly, the indifference curves too will be linear, and if it is convex, the indifference curves will be concave. Finally, if voter loss functions follow the reverse S-shaped function, the indifference curves will be convex at high values of utility and for low values, concave. So although the voter indifference curves differ between the two approaches across each of the functional form assumptions, the general trends regarding the separation between the indifference curves do not. Since voter indifference is best reflected through the separation between indifference curves, the functional form assumption that best predicts voter indifference will be the same in both the issue and spatial based approaches. This opens new questions as to whether voters' indifference curves are convex across the policy space. If it is truly the case that voter's preference relations are best captured by the issue based models and reverse S-shaped loss functions, voter preferences are expected not to be always convex. This possibility also suggested by Kamada and Kojima (2012), bears for an answer which we hope will be investigated in future research.

Finally, our findings also have implications for the directional approach to studying voter behavior. With directional voting, voter utility from a given candidate is modeled as a function of the directional alignment of their preferences and the extremity of candidate position in these directions, or $\vec{x_i}\cdot \vec{x_c}= |\vec{x_i}||\vec{x_c}|cos\theta_{i,c}$ where $\theta_{i,c}$ is the angle between position vectors $\vec{i}$ and $\vec{c}$ . Even though the second term of this formulation, $cos\theta_{{i},{c}}$, establishes a mechanism reminiscent of the reverse S-shaped functional form assumption, the first term generates predictions implying that the voters grow less indifferent to candidates in more polarized races, similar to the concave loss function predictions. That is, as the candidates become more polarized as the term $|\vec{x_i}||\vec{x_c}|$ will be increasing in candidate polarization, the voter's indifference to the candidates is expected to decrease. However, especially in \emph{Case 2}, it is exhibited that this prediction does not match our empirical observations.

\begin{center}
\begin{table}
\caption{\normalsize Functional Form Assumption's Implications for Voter Indifference under Spatial and Issue-Based Models}

\fontsize{10}{12}\selectfont
\label{fig_indiff_curves}

     \begin{tabular}{|c|l|c|} 
        \hline 
         &Spatial Model&
        Issue-Based Model\\
        \hline 
        Linear & \tikzpicone 
&
        \tikzpicfive\\ 
        \hline 
        Strictly Concave & \tikzpictwo 
&
        \tikzpicsix\\ 
        \hline
        Strictly Convex & \tikzpicthree 
&
        \tikzpicseven\\
        \hline
        Reverse S-shaped & \tikzpicfour &
        \tikzpiceight\\
        \hline
    \end{tabular}
\end{table}

\end{center}

\clearpage

\setcounter{table}{0}
\setcounter{figure}{0}

\section{Additional Support: Tables and Figures}

\begin{center}
\begin{table}[!h]

\caption{Proportion of Voter Group Subgroup Pairs with Downward Change in Democratic Support}
\centering
\begin{tabular}[t]{lr}
\toprule
\textbf{Proposition} & \textbf{Proportion}\\ [.05in]
 \multicolumn{2}{c}{\textit{California}}\\ [.05in]
\midrule
Proposition 14 & 0.980\\
Proposition 15 & 0.994\\
Proposition 16 & 0.994\\
Proposition 17 & 1.000\\
Proposition 18 & 1.000\\
Proposition 20 & 0.994\\
Proposition 21 & 0.840\\
Proposition 22 & 0.990\\
Proposition 24 & 0.195\\
Proposition 25 & 0.982\\ [.05in]
 \multicolumn{2}{c}{\textit{Colorado}}\\ [.05in]
\midrule
 Amendment 76 & 1.000\\
 Amendment B & 0.926\\
 Proposition 113 & 1.000\\
 Proposition 114 & 0.992\\
 Proposition 115 & 1.000\\
 Proposition 116 & 1.000\\
 Proposition 117 & 0.852\\
 Proposition 118 & 1.000\\
 Proposition EE & 0.988\\
 \bottomrule
\end{tabular}
\label{tab_prop_down} 

\end{table}

\end{center}

\begin{figure}[h!]
\centering
\begin{minipage}{0.42\textwidth}
    \centering
    \includegraphics[width=\textwidth]{figures/posdist_DemVShare_Bar_CA.pdf}
\end{minipage}
\begin{minipage}{0.42\textwidth}
    \centering
    \includegraphics[width=\textwidth]{figures/posdist_DemVShare_Bar_CO.pdf}
\end{minipage}
\begin{minipage}{0.42\textwidth}
    \centering
    \includegraphics[width=\textwidth]{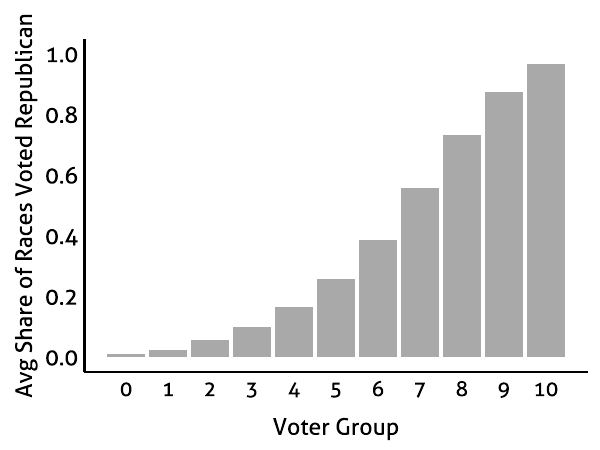}
\end{minipage}
\begin{minipage}{0.42\textwidth}
    \centering
    \includegraphics[width=\textwidth]{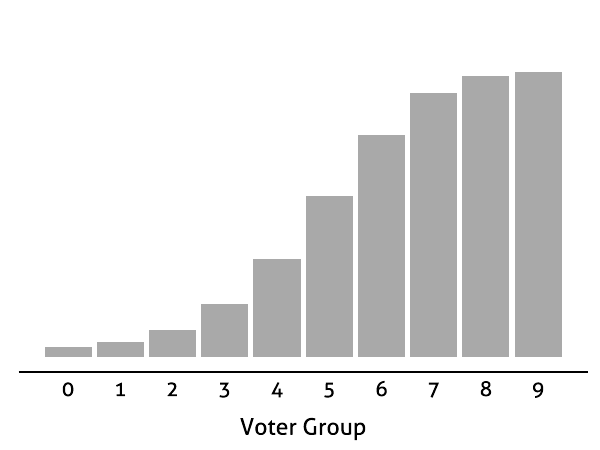}
\end{minipage}
\vspace{-0.2cm}
\caption{Average Share of Races Voted Democrat (Top Panel) and Republican (Bottom Panel) by \emph{Voter Group} for California (left) and Colorado (right)}
\label{figbar_VGroup_DemRepVShare_append}
\end{figure}

\begin{figure}[h!]
\begin{center}
 \includegraphics[scale=0.98]{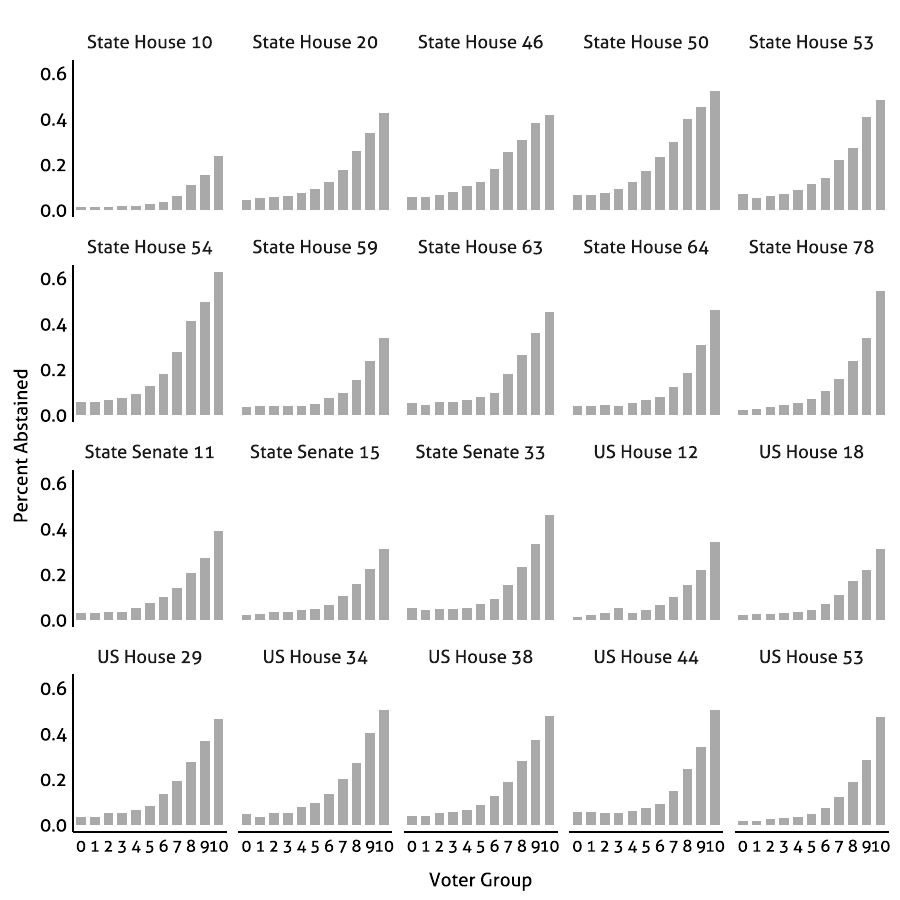}
\end{center}
\vspace{-0.45cm}
 \caption{\textbf{Abstention in Democrat v. Democrat Races, 2020 General Election California}}
 \label{fig_dd_roll_race}
 \vspace{-0.2cm}
\singlespacing
\small
 \textit{Note}: Barplots display the percent of individuals who abstained in the House and State Assembly races where a Democrat faced a Democrat for each value of \emph{Voter Group}.  In 2020, seven US House races - District's 12, 18, 29, 34, 38, 44, and 53 -  eleven State House races  - District's 10, 13, 20, 46, 50, 53, 54, 59, 63, 64, 78, and three State Senate races - District's 11, 15, and 33 - featured two Democratic candidates.  Please see the Data Description section for complete details on the \emph{Voter Group} measure, including the ballot measure language, partisan support, and complete distribution.  
\end{figure}

\begin{figure}[h!]
\begin{center}
 \includegraphics[scale=0.98]{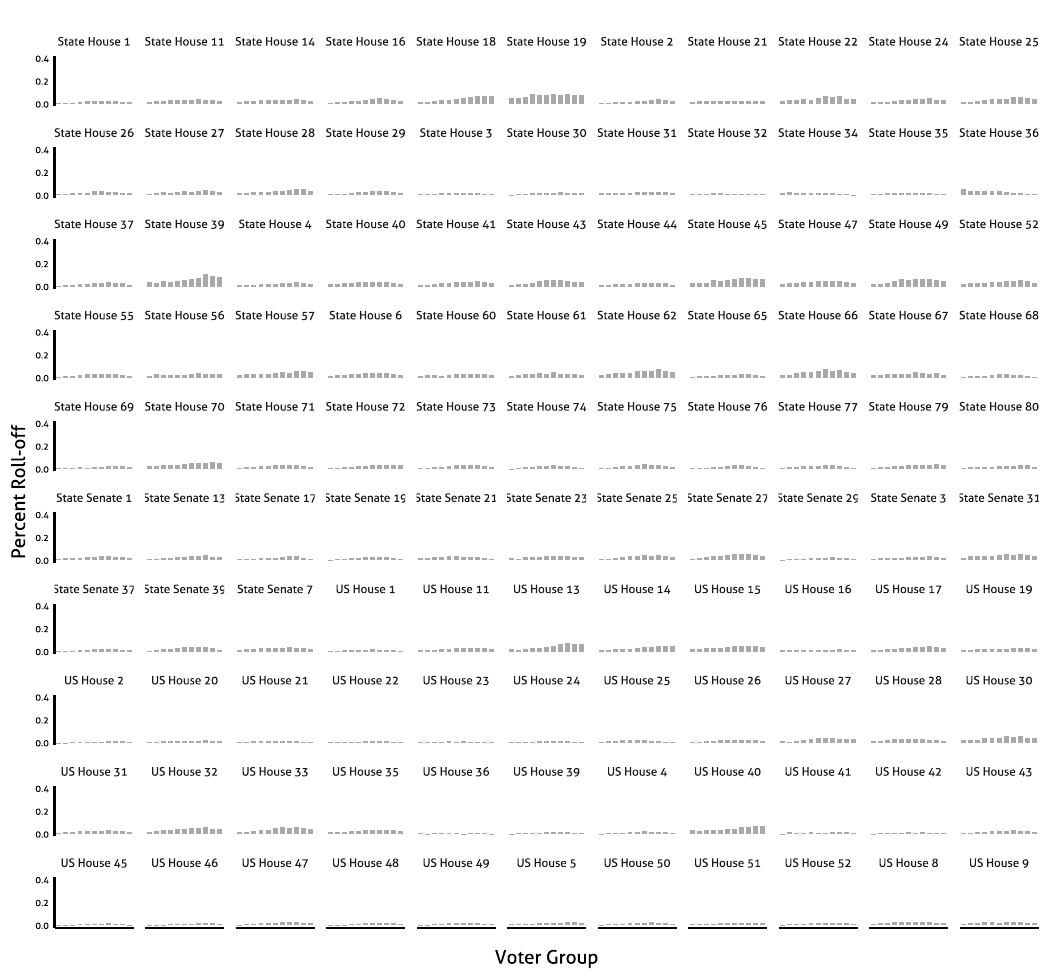}
\end{center}
\vspace{-0.45cm}
 \caption{\textbf{Abstention in Democrat v. Republican Races, 2020 General Election California}}
 \label{fig_DR_roll_race}
 \vspace{-0.2cm}
\singlespacing
\small
 \textit{Note}: Barplots display the percent of individuals who abstained in the House and State Assembly races where a Democrat faced a Republican for each value of \emph{Voter Group}.  Please see the Data Description section for complete details on the \emph{Voter Group} measure, including the ballot measure language, partisan support, and complete distribution.  
\end{figure}

\begin{figure}[h!]
\begin{center}
 \includegraphics[scale=0.80]{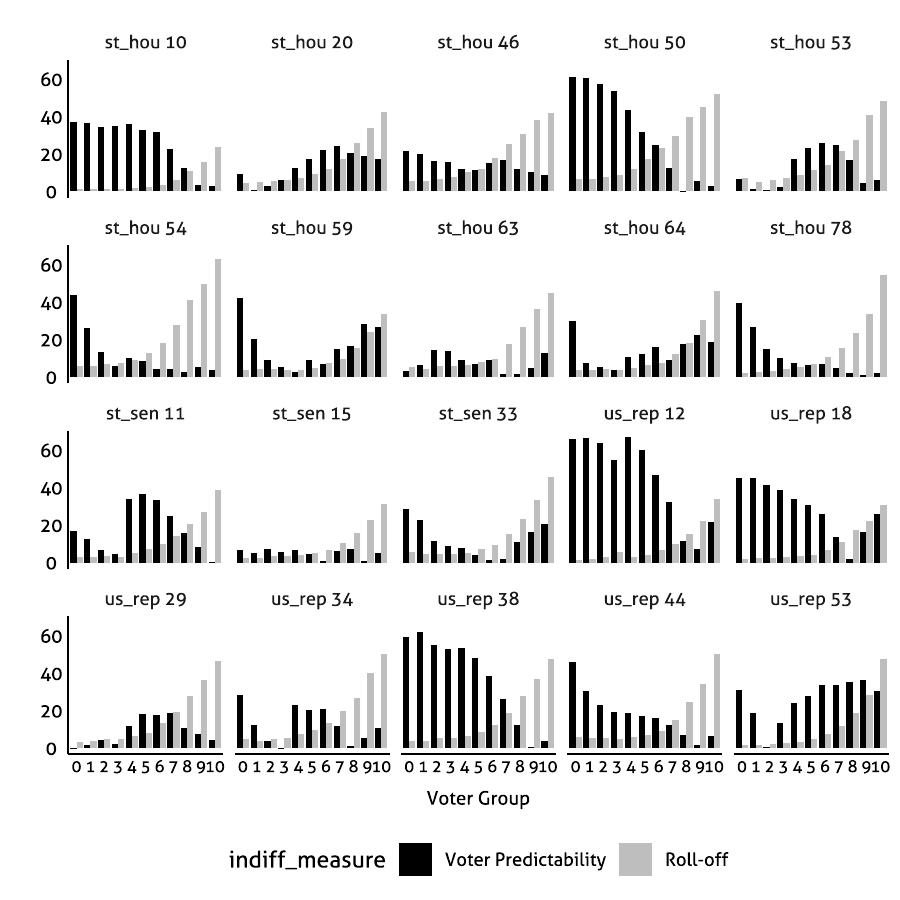}
\end{center}
 \vspace{-0.45cm}
 \caption{\textbf{Comparing Two Measures of Average Voter Group Indifference Within All Races Between Candidates From the Same Party}}
\label{fig_indiff_barplot_race}
\end{figure}

\begin{figure}[h!]
\begin{center}
 \includegraphics[scale=0.50]{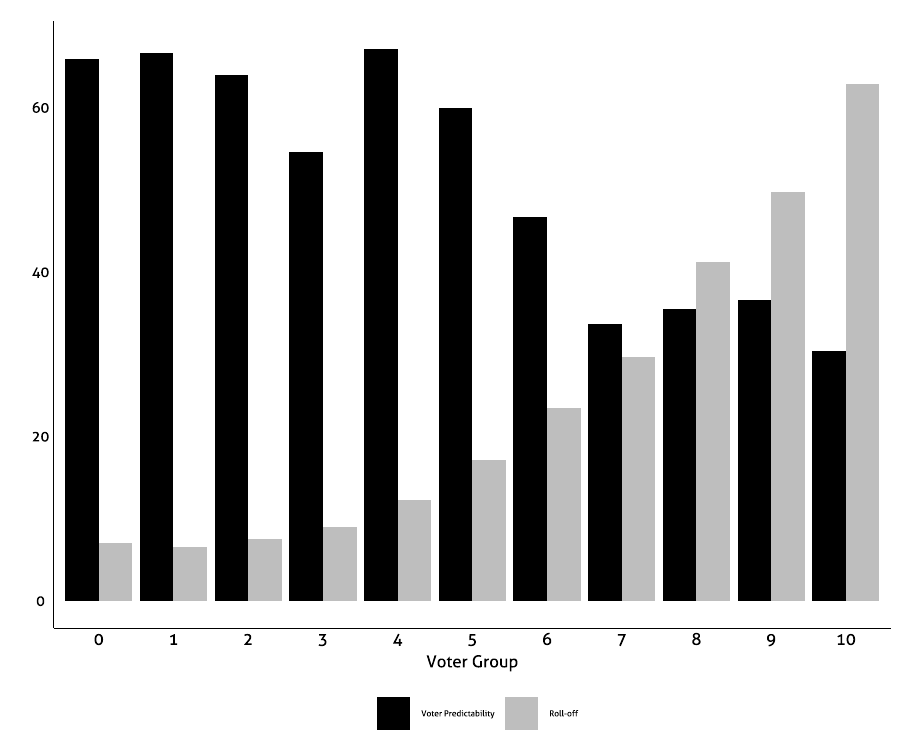}
\end{center}
 \vspace{-0.45cm}
 \caption{\textbf{Comparing Two Measures of Average Voter Group Indifference Across All Races Between Candidates From the Same Party}}
\label{fig_indiff_barplot_all}
\end{figure}

\begin{figure}[h!]
\begin{center}
 \includegraphics[scale=0.80]{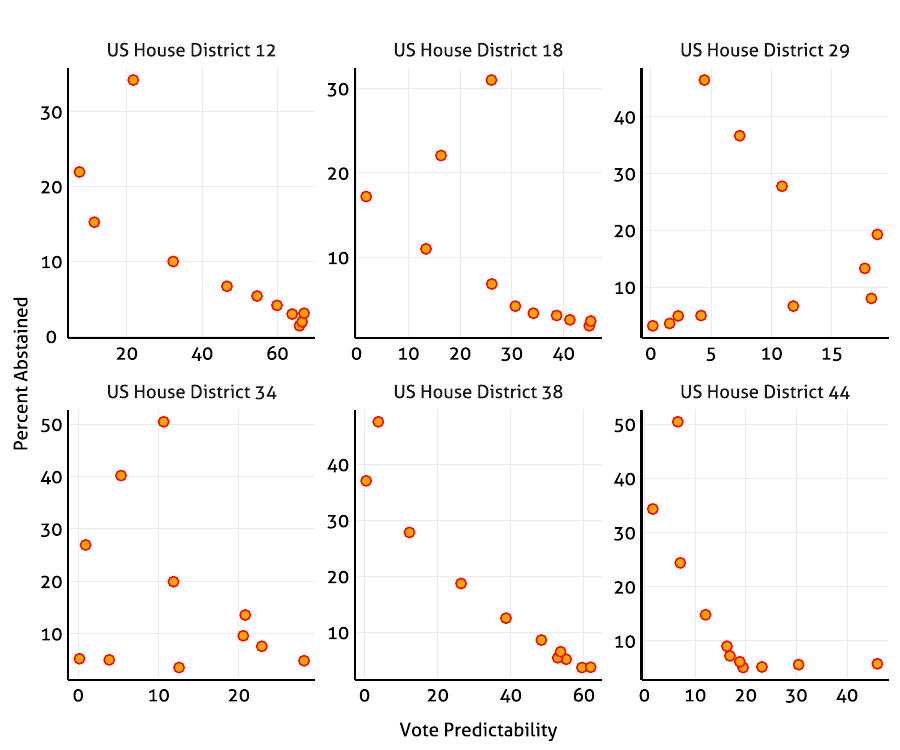}
\end{center}
 \vspace{-0.45cm}
 \caption{\textbf{Correlation Between Two Measures of Voter Indifference in California Democrat vs. Democrat US House Races: Abstention and Voter Predictability}}
\label{fig_indiff_corr_rep}
\end{figure}

\newpage
\clearpage
\begin{figure}[h!]
\begin{center}
 \includegraphics[scale=0.99]{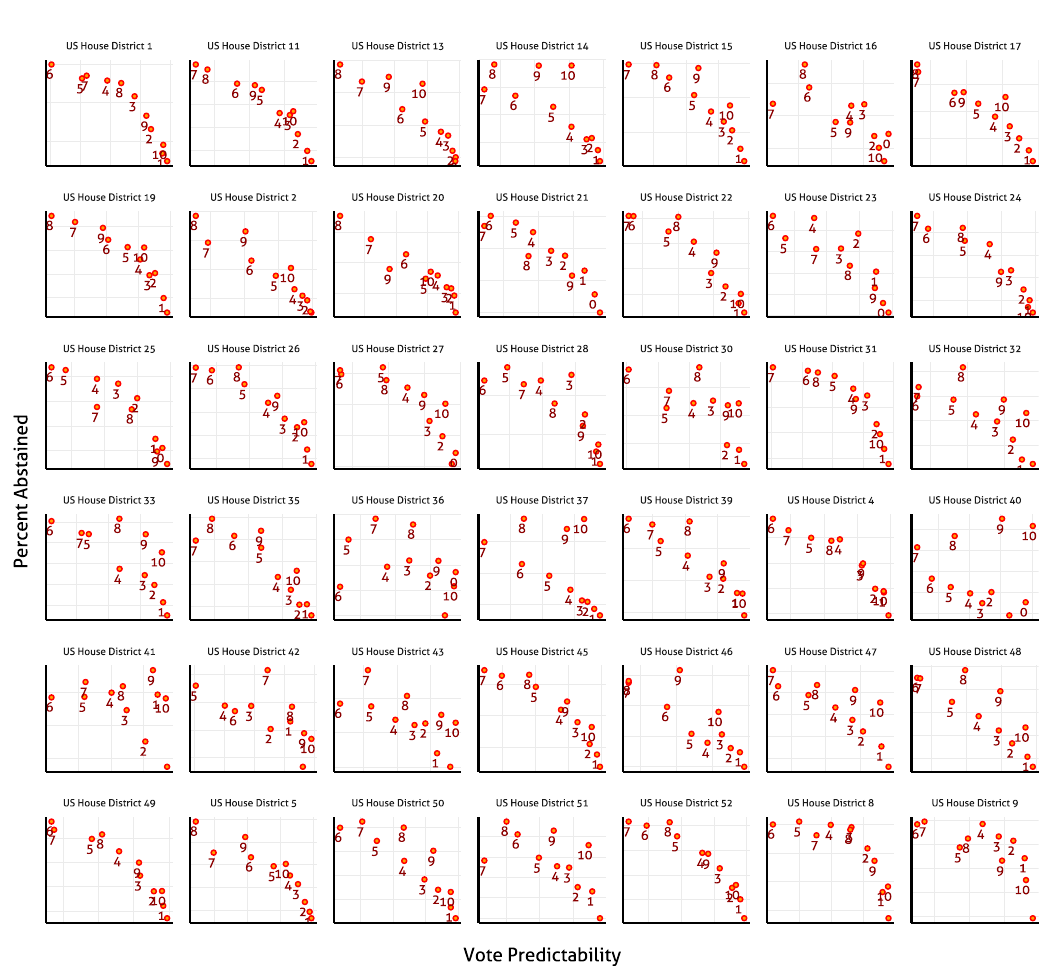}
 \end{center}
 \vspace{-0.45cm}
 \caption{\textbf{Correlation Between Two Measures of Voter Indifference in California Democrat vs. Republican US House Races: Abstention and Voter Predictability}}
\label{fig_indiff_corr_rep_CA}
\end{figure}

\newpage
\clearpage
\begin{figure}[h!]
 \includegraphics[scale=0.99]{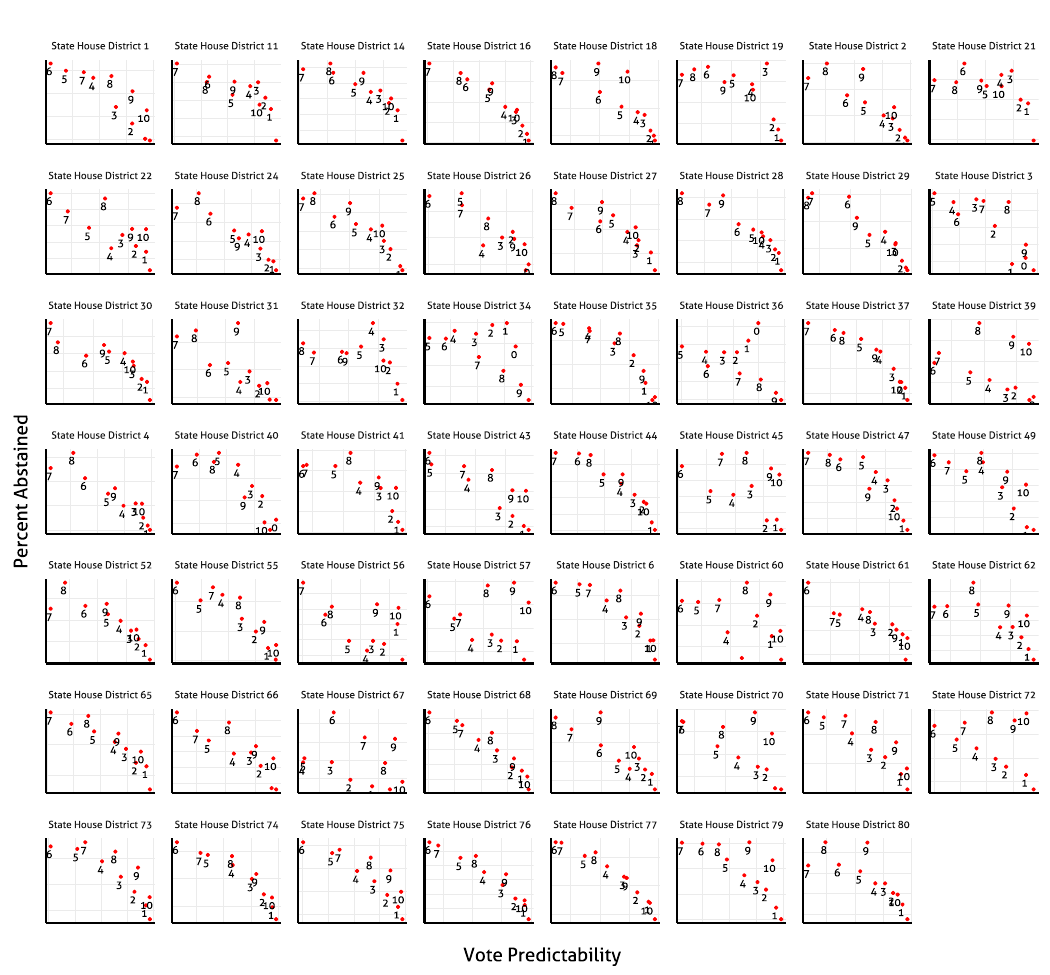}
 \vspace{-0.45cm}
 \caption{\textbf{Correlation Between Two Measures of Voter Indifference in California Democrat vs. Republican State House Races: Abstention and Voter Predictability}}
\label{fig_indiff_corr_StHou_CA}
\end{figure}

\newpage
\clearpage
\begin{figure}[h!]
\centering
 \includegraphics[scale=0.99]{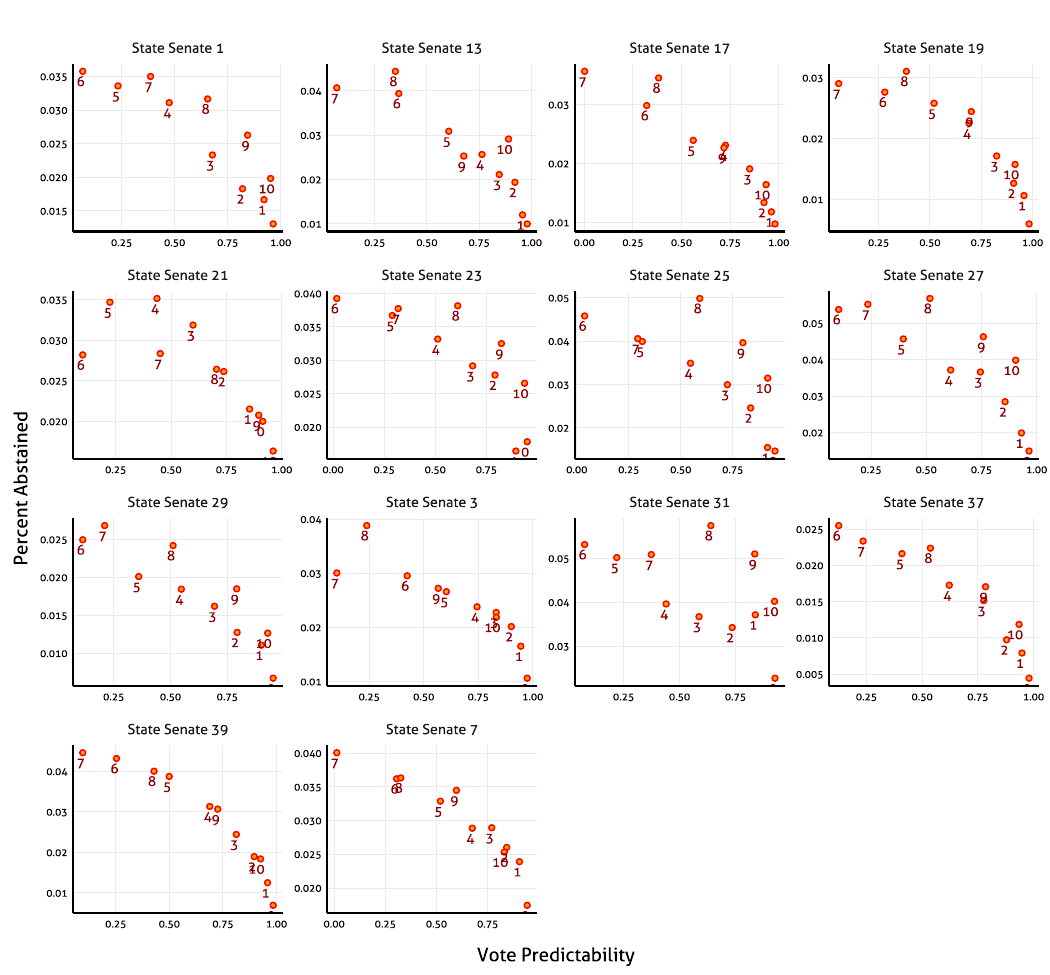}
 \vspace{-0.45cm}
 \caption{\textbf{Correlation Between Two Measures of Voter Indifference in California Democrat vs. Republican State Senate Races: Abstention and Voter Predictability}}
\label{fig_indiff_corr_StSen_CA}
\end{figure}

\begin{figure}[h!]
 \includegraphics[scale=0.99]{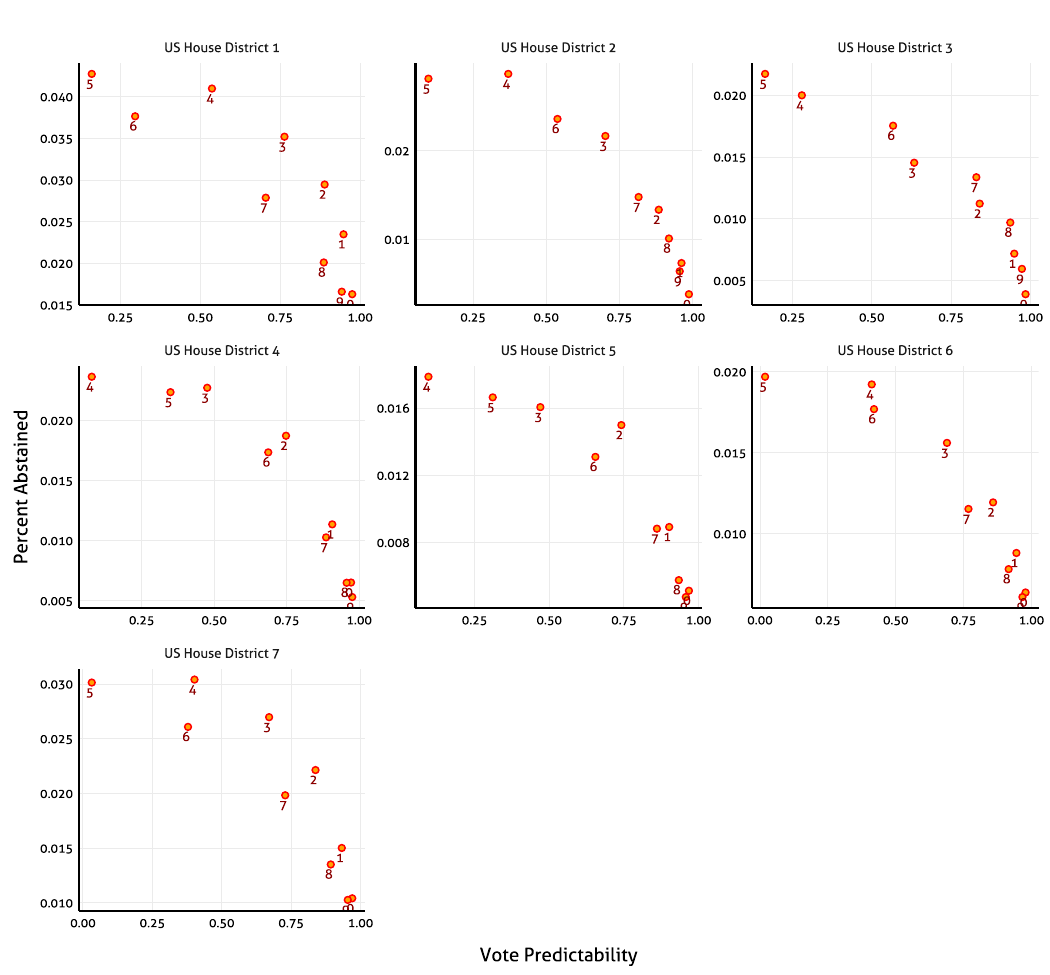}
 \vspace{-0.45cm}
 \caption{\textbf{Correlation Between Two Measures of Voter Indifference in Colorado Democrat vs. Republican US House Races: Abstention and Voter Predictability}}
\label{fig_indiff_corr_rep_CO}
\end{figure}

\begin{figure}[h!]
\begin{center}
 \includegraphics[scale=0.99]{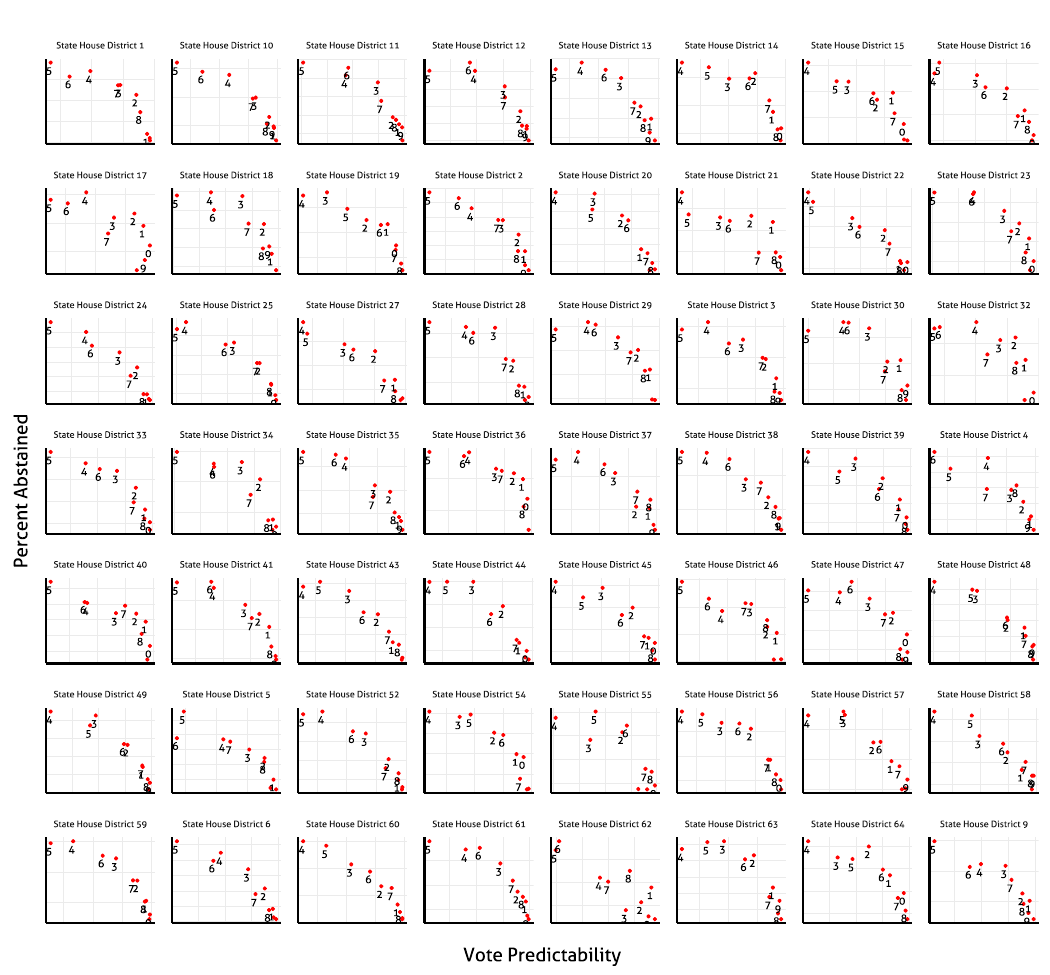}
\end{center}
 \vspace{-0.45cm}
 \caption{\textbf{Correlation Between Two Measures of Voter Indifference in Colorado Democrat vs. Republican State House Races: Abstention and Voter Predictability}}
\label{fig_indiff_corr_StHou_CO}
\end{figure}

\begin{figure}[h!]
\begin{center}
 \includegraphics[scale=0.99]{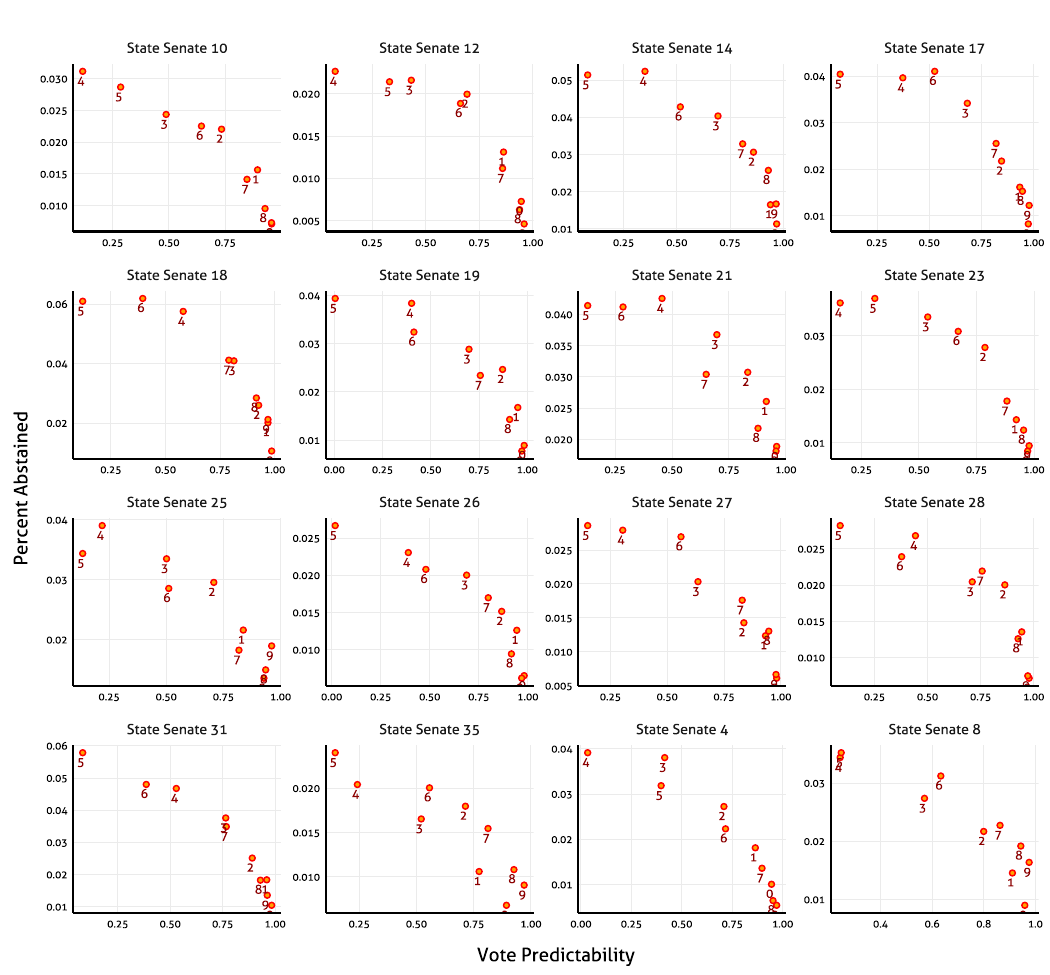}
\end{center}
 \vspace{-0.45cm}
 \caption{\textbf{Correlation Between Two Measures of Voter Indifference in Colorado Democrat vs. Republican State Senate Races: Abstention and Voter Predictability}}
\label{fig_indiff_corr_StSen_CO}
\end{figure}

\begin{figure}[h!]
\begin{center}
 \includegraphics[scale=0.6]{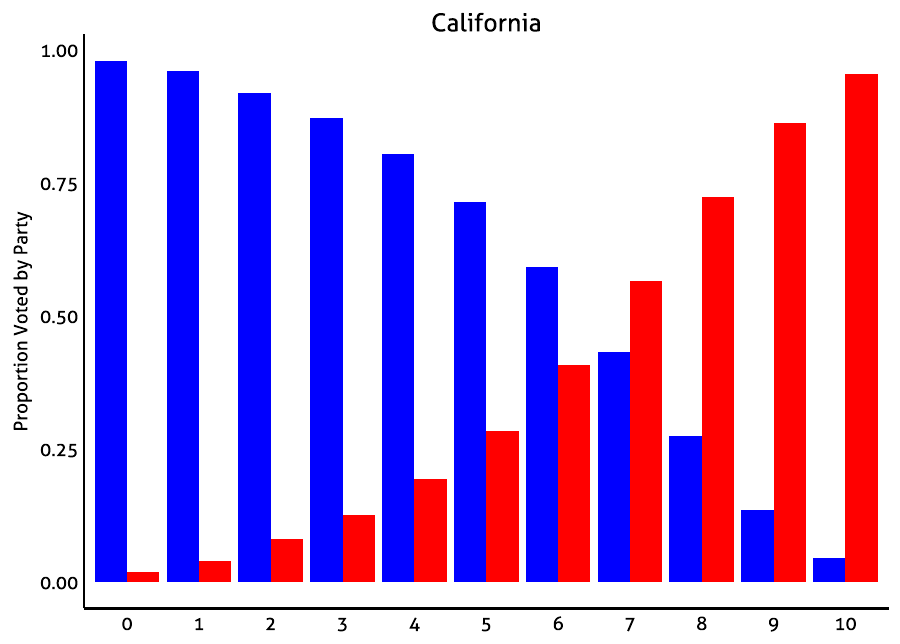}
 \includegraphics[scale=0.6]{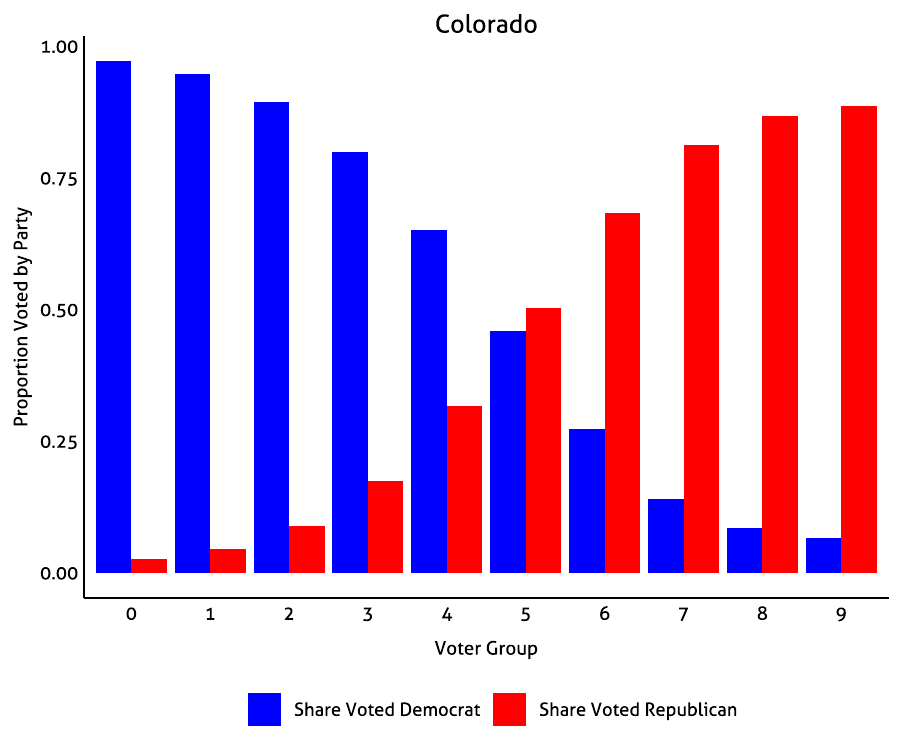}
\end{center}
 \vspace{-0.45cm}
 \caption{\textbf{Average Proportion of Support for the Two Major Parties Across National and State Level Races by Voter Group}}
\label{fig_moderates_ca_co}
\end{figure}

\newpage
\clearpage
\begin{center}
\begin{table}[!h]
\centering
\caption{R-Squares From Regressions of Voter Group, PCA on Vote Choice}
\fontsize{10}{12}\selectfont
\label{fig_indif_wrace}
\begin{tabular}[t]{lccrl}
\toprule
State & Office & Voter Group & PCA & Obs\\
\midrule
CA & US President & 0.498 & 0.475 & 7,651,675\\
CA & US House & 0.508 & 0.487 & 6,763,822\\
CA & State House & 0.497 & 0.482 & 5,817,728\\
CA & State Senate & 0.525 & 0.508 & 3,202,223\\
CO & US President & 0.439 & 0.455 & 2,675,387\\
CO & US Senate & 0.436 & 0.453 & 2,675,387\\
CO & US House & 0.536 & 0.557 & 2,675,400\\
CO & State House & 0.538 & 0.559 & 2,396,781\\
CO & State Senate & 0.537 & 0.559 & 1,243,485\\
\bottomrule
\end{tabular}
\end{table}

\end{center}

\begin{center}

\begin{table}[!h]
\centering
\caption{Correlation Between Indifference Measures Within Voter Group, 
  Colorado Democrat vs. Republican Races}
\fontsize{11}{13}\selectfont
\label{corr_indiff_table_CO}
\begin{tabular}[t]{lc}
\toprule
\multicolumn{1}{c}{ } & \multicolumn{1}{c}{Abstention} \\
\cmidrule(l{3pt}r{3pt}){2-2}
  & (1)\\
\midrule
Vote Predictability & \num{-0.011}*\\
 & (\num{0.004})\\
\midrule
No. of Observations & \num{810}\\
Voter Group Fixed Effects & Yes\\
\bottomrule
\multicolumn{2}{l}{\rule{0pt}{1em}\textit{Note: } Standard errors in parentheses.}\\
\multicolumn{2}{l}{\rule{0pt}{1em}* p $<$ 0.01}\\
\end{tabular}
\end{table}

\end{center}

\begin{center}
\begin{table}[!h]
\centering
\caption{Correlation Between Indifference Measures Within Voter Group, 
  California Democrat vs. Republican Races}
\fontsize{11}{13}\selectfont
\label{corr_indiff_table_CA}
\begin{tabular}[t]{lc}
\toprule
\multicolumn{1}{c}{ } & \multicolumn{1}{c}{Abstention} \\
\cmidrule(l{3pt}r{3pt}){2-2}
  & (1)\\
\midrule
Vote Predictability & \num{-0.022}*\\
 & (\num{0.003})\\
\midrule
No. of Observations & \num{1243}\\
Voter Group Fixed Effects & Yes\\
\bottomrule
\multicolumn{2}{l}{\rule{0pt}{1em}\textit{Note: } Standard errors in parentheses.}\\
\multicolumn{2}{l}{\rule{0pt}{1em}* p $<$ 0.01}\\
\end{tabular}
\end{table}

\end{center}

\end{document}